\documentstyle[11pt,aaspp4]{article}

\def\vs{\vskip 0.2cm}

\def\->{$\longrightarrow$}

\def\ie{{\it i.e.,~}}

\def\h0{H_0}
\def\q0{q_0}

\def\acf{\omega(\theta)}

\def\peak{I_{peak}}
\def\npix{N_{pix}}
\def\cin{C_{in}} 

\newcommand{\beq}{\begin{equation}}
\newcommand{\eeq}{\end{equation}}
\newcommand{\barr}{\begin{eqnarray}}
\newcommand{\earr}{\end{eqnarray}}

\received{}
\accepted{}

\lefthead{Kashikawa et al.}
\righthead{The Morhology Dependence of Luminosity Segregation in the Coma Cluster}

\begin{document}

\title{The Morphology Dependence of Luminosity Segregation in the Coma Cluster}

\author{Nobunari Kashikawa and Maki Sekiguchi\altaffilmark{1}}
\affil{Optical and Infrared Astronomy Division, National
Astronomical Observatory of Japan, Mitaka, Tokyo 181, Japan}


\and

\author{Mamoru Doi, Yutaka Komiyama, Sadanori Okamura, Kazuhiro Shimasaku, Masafumi Yagi, and Naoki Yasuda}
\affil{School of Science, University of Tokyo, Bunkyo-ku, Tokyo 113, Japan}


\altaffiltext{1}{Present address: Institute for Cosmic Ray Research,
University of Tokyo, Tanashi, Tokyo 188} 


\begin{abstract}
We carry out CCD photometry of galaxies in the $5.25$ square region 
centered on Coma cluster down to $M_R=-16.0$, beyond the limit
of conventional morphological classification.
We use the angular two-point correlation function as well as
radial profiles in order to characterize the luminosity segregation.
We find strong luminosity segregation for our total sample over the
magnitude range of $-20 \leq M_R \leq -16$, which is not entirely 
accounted for in terms of the morphology-density relation that is known 
to exist only for bright galaxies.
We use a single consistent parameter, the degree of luminosity 
concentration, to parameterize the morphology of galaxies over
the wide magnitude range, where both giant and dwarf galaxies are
included. Galaxies with high central concentration (HCC) show strong 
luminosity segregation, \ie their clustering strength depends strongly
on luminosity while those with low central concentration (LCC)
show almost no luminosity segregation.
Radial density profile shows that brighter HCC-type galaxies tend to more 
strongly concentrate near the cluster center while LCC-type galaxies do 
not show such a dependence on luminosity.
We show that these results are tenable against the contamination
by field galaxies and uncertainties in our method of classification.
\end{abstract}




\section{INTRODUCTION}

Some studies suggest that luminous galaxies are clustered more
strongly than faint galaxies. 
This phenomenon is referred to as the luminosity segregation (hereafter LS). 
The LS can be interpreted as a result of either primordial effects or environmental effects.
Explanations based on the primordial effect include the biased Cold
Dark Matter (CDM) model in which galaxies form at high peaks in the density field (e.g., \cite{kai84}; \cite{dav85}).
It is known that the LS is naturally predicted by this model.
Valls-Gabaud, Alimi, \& Blanchard (1989) pointed out that the
correlation strength has a positive dependence on galaxy luminosity on
the basis of the biased-CDM model.
White et al. (1987) also predicted that clustering strength is a strong function of the circular velocity of galaxies, which in turn correlates with the luminosity as indicated by Tully-Fisher or Faber-Jackson relations.
In terms of environmental effects, the LS can be regarded as a result
of frequent merging or other dynamical mechanisms in the vicinity of the cluster core.

Observational evidence for the LS has been uncertain and controversial.
Some found positive results (e.g., Capelato et al. 1980;
Dom\'inguez-Tenreiro \& Pozo-Sanz 1988; Davis et al. 1988), and
others found negative results (e.g., Phillips \& Shanks 1987; Einasto 1991).
The clustering property of galaxies is also correlated with their
morphology, which is so-called the morphology segregation, or
morphology-density relation (\cite{dre80b}).
If the morphology segregation is the fundamental correlation, the LS
would be naturally expected.  
Early-type galaxies show a stronger degree of clustering than
late-type galaxies, and at the same time, early-type galaxies are on
the average brighter than late-type galaxies (e.g., \cite{eep88}). 
Consequently, galaxies which are more strongly clustered are brighter
than those which are less clustered, which is the LS. 
On the contrary, if the LS is an essential correlation, the
morphology segregation would also be expected. 
Accordingly, it is critical to see if the LS is observed {\it within a given
morphological type} in order to disentangle the coupling of the LS and
morphology segregation. 
Only a few such studies have been made so far (\cite{ein91}; \cite{lov95}).

There are also few studies on the LS among dwarf galaxies.
Binggeli, Tammann, \& Sandage ($1987$) found in the Virgo cluster that
nucleated dwarf ellipticals (dEs) are more strongly concentrated towards the cluster
center than nonnucleated dEs.
Ferguson \& Sandage ($1989$) found in the Virgo and Fornax clusters that
the faint ($M_B>-13.3$) nonnucleated dEs have the distribution
identical to that of the E/S0 galaxies and bright nucleated dEs.
They also found that the faint nonnucleated dEs are more strongly 
concentrated on the cluster center than the bright nonnucleated dEs.
Thompson \& Gregory ($1993$) showed in the Coma cluster that both
dEs and dwarf spheroidals have the same distributions as 
that of giant early-type galaxies.
Morphological classifications of these studies are based on the
eye inspection and Thompson \& Gregory's criteria used for Coma dwarfs 
are slightly different from those of Sandage and collaborators used for Virgo dwarfs.

One of the reasons why the LS has not been examined systematically is the difficulty in sampling a large number of faint galaxies with known absolute magnitude. 
Clusters of galaxies are good targets to address the problem of the LS in high density environments.
In  particular, in nearby clusters we can sample intrinsically faint galaxies, which are critically important to the study of the LS. However, nearby clusters have such a large apparent size that we cannot survey whole the cluster with a CCD which has a small physical size.

In this study, we present the angular two-point correlation function
as well as the radial profiles of galaxies in the Coma cluster on the basis of a large homogeneous sample covering a wide magnitude range
$-20 \leq M_R \leq -16$, where both giant and dwarf galaxies are
included.
We examine if there is a difference in these properties between the galaxies 
which have high central concentration of the surface brightness
distribution and those with low central concentration.
Our sample is made available by three new techniques; CCD
mosaic, semi-automated data reduction/analysis software, and
quantitative and objective classification of morphological type of
galaxies based on surface photometry parameters.

In section $2$, we briefly explain our imaging observation of the Coma cluster.
In section $3$, we describe the data reduction procedures, calibration methods which are special to our camera, and the method of constructing a homogeneous galaxy sample.
In section $4$, we describe star-galaxy discrimination, evaluation of
the number of contaminated field galaxies, and the method of
classifying galaxies according to the surface brightness
concentration.
We present our results in section $5$. 
Finally, reliability of our results is discussed in section $6$ in terms of the effects of uncertainties in our classification and contamination by background galaxies.
A comparison with previous studies are given in section $7$. 

\section{OBSERVATIONS}

We observed the Coma cluster region on May $1$--$2$, $1992$ and March
$17$--$21$, $1993$ with our mosaic CCD camera (hereafter MCCD1)
mounted on the prime focus of the Kiso $105$cm Schmidt telescope.
MCCD1 consists of an $8\times2$ array of $1000\times1018$-pixel CCDs.
CCDs are placed with a relatively large space between the chips (Figure~\ref{fcfg1}).
One contiguous field is completed by taking several exposures, each shifted by a fixed amount on the CCD grid.
A technical description of MCCD1 is presented in Sekiguchi et al. ($1992$).

The image scale at the Kiso Schmidt prime focus is $0.^{\prime\prime}75$/pixel.
The sky coverage is $1.^\circ7\times3.^\circ4=5.25$ square degrees and the center of the surveyed region is ($12^h 59^m 35.^s2$, $+27^{\circ} 57^{\prime} 35^{\prime\prime}$) (J$2000.0$).
One exposure of our camera is referred to as a ^^ shot', which produces $16$ frames for MCCD1.
We have to take $15$ shots for MCCD1 to complete the survey of one contiguous field.
A contiguous image consists of about $16250\times 8100$ pixels constructed from $16\times15=240$ frames.
One CCD chip was dead, and thus the data are lacking in the southwest corner ($\sim0.35$ square degrees).
We obtained $225$ frames ($15$ chips times $15$ shots) of a $20$ minutes exposure in the $R$ band.
The seeing size was $3.^{\prime\prime}8$ FWHM on average. 

\placefigure{fcfg1}

\section{DATA REDUCTION PROCEDURES}

The data produced by our mosaic CCD camera amount to about $1$ to $3$ GB per night.
We have developed a suit of software which reduces these large amount of data.
Details of this software package will be given by Yagi et al. (1998).
The best performance of the package is attained by two-pass processing.
The first pass is to fine-tune several critical parameters that are used in the second pass, which produces the final catalog of objects with photometric parameters together with parameters necessary to construct one contiguous image.

Functions of our software package contain 
(1) flat fielding, (2) sky subtraction, (3) bright objects detection,
(4) mosaicking, (5) smoothing, (6) objects detection and
parameterization, (7) photometric calibration, (8) astrometric calibration, (9) star/galaxy discrimination, and (10) morphological classification.
The tasks (1)-(6) are carried out ^^ on-line', \ie in a semi-automatic
routine basis while tasks (7)-(10) are performed ^^ off-line'.
Tasks (1)-(8) are briefly explained below and tasks (9) and (10) are described in section $4$.

\subsection{ Flat Fielding and Sky Subtraction} \label{hairymath}

In flat fielding, each frame is reduced according to the ordinary way of CCD data reduction.
The sky-flat frame for each chip is constructed by taking the median value for each pixel of a lot of object frames ($15$ -- $20$ frames) exposed with the chip at different sky regions.
We also construct the median dome-flat frame as well as the sky-flat frame.
The flat frame is constructed by multiplying the sky-flat frame with the dome-flat frame and normalizing the resultant frame so that the average count of all the pixels becomes unity.
Bias subtraction and flat fielding, \ie dividing by the flat frame, are carried out independently for each CCD chip.
Dark current was negligible and no fringe or interferometric pattern was seen in our data.

We subtract the sky-background count from the flat fielded frames.
The local sky background is computed as follows.
First, we divide each frame into $10\times10$ meshes, each consisting of $100\times100$ pixels.
A histogram of pixel counts is constructed for every mesh.
The peak of the histogram is determined by fitting a Gaussian to the low count side of the histogram and the count at the peak of the Gaussian is regarded as the representative sky count for the mesh.
The representative sky counts are assigned to the central pixels of respective meshes.
A bi-linear interpolation of those representative counts is performed to determine the local background sky count of every pixel in the frame, which is subtracted from the pixel count of the flat fielded frame.

\subsection{Coarse Finding of Bright Stars}

Before we detect objects and measure their photometric parameters, the coarse detection is done to find some parameters for constructing a homogeneous sample all over the survey field.
Our object finding is based on the so-called ^^ connected pixel method'.
We recognize a lump of pixels whose counts are above a threshold value, $I_{th}$, as an object when it consists of more pixels than a minimum number, $N_{min}$.
We empirically set the minimum number to the number of pixels enclosed within the circle with a diameter of the FWHM of the seeing size, and the threshold value to be several times of sky  r.m.s. fluctuations above the sky.
The detected objects in this procedure are used for seeing measurement and mosaicking which will be explained below.

\subsection{Seeing Measurement}

In each frame, images of typical stars are scissored out and shifted so that the centers of gravity of the stellar images are coincident.
Then, all the images are added together to give a combined stellar 
image.
We measure the seeing size by fitting a Gaussian to the central region of the combined stellar image.
We take the FWHM of the Gaussian as the seeing size.
The seeing size of frames exposed on different chips should be, in principle, identical for the same exposure.
We find a small change in the seeing size for different frames even within the same exposure.
This may be due to the slight tilt of the dewar mount with respect to the optical axis.
A tiny distortion of the image caused by guide errors is also found in
this combined stellar image.
In this paper, we refer the ^^ seeing size' to the FWHM size of the
Gaussian fitted to this combined stellar image. 

\subsection{Smoothing}

Now we have obtained the seeing size for all the frames. 
The difference in the seeing size among different exposures, if
significant, should be corrected for by smoothing the data at least to the first order approximation.
The frames with smaller seeing sizes should be smoothed with a Gaussian beam so that all the frames for a contiguous field have the same seeing size.
This seeing equalization procedure is important to obtain a homogeneous photometric sample over the field because we derive isophotal magnitudes and sizes which are affected easily by seeing, especially, in case of faint galaxies (\cite{yos93}).

We finally adjust seeing size at $\sim4.^{\prime\prime}5$. 
In spite of the seeing equalization, there is a little residual tendency that the chips on south-side (larger number in each raw) show larger FWHM.
This is supposed to be due to an asymmetric distortion of star images
caused by a slight tilt of dewar mounting.

\subsection{Mosaicking}

The most important process, that is, the frame mosaicking, is attained
on the basis of the position and the flux of the stars in the overlapped region between adjacent frames.
This process is important because our goal is to obtain a statistically homogeneous sample of galaxies over a wide area detected by the same criteria and measured in the same manner using a mosaicked image which consists of a lot of frames taken at different nights
 and under different observational conditions.

In the first step of this procedure, triangles of stars are constructed in the overlapped region.
A triangle is composed of three stars.
Triangles are made for all possible combinations of three stars extracted from the detected stars in the overlapped region.
Every triangle on a frame is paired with the corresponding triangle on the adjacent frame to be mosaicked.
A pair of the same stars in the two frames is identified by this triangle pattern-matching algorithm (\cite{gro86}).
We determined the size of the overlapped region on the basis of the star count at the north galactic pole (\cite{bs83}, hereafter BS83) so that enough stars are included.
In fact, at least three pairs of stars were obtained in any overlapped region.  
In the second step, adjacent frames are shifted, rotated and flux-scaled as follows so that the position and the flux of these star pairs coincide with each other in both frames.
The algorithm used in this process is contrived so that the errors in position and flux are evenly distributed over the whole set of frames.
We determine four parameters necessary to the mosaicking, \ie x, y positions, rotation angle, and flux scaling factor of each frame by minimizing the r.m.s. error over the whole set of frames under the restriction that the cumulative sum over any possible loops involving neighboring $2\times2$ frames would be zero (see Yagi et al. 1998).

The mosaicking error becomes larger as magnitude goes fainter.
The r.m.s. of mosaicking errors in positions is estimated to be $0.17$ and $0.21$ pixels in the magnitude range $17.5 < R \leq 18.0$ for x and y, respectively.
The r.m.s. of mosaicking errors in magnitude at the faintest bin $17.5 < R \leq 18.0$ is $0.08$mag.

\subsection{Finding Objects and Parameterization}

Finally the software package outputs a catalog including all the detected objects.
We determined the threshold level $I_{th}$ to be $1.5$ times the largest sky
r.m.s. noise among the frames. 
The minimum number $N_{min}$ was determined to the number of pixels within
the circle with a diameter of the equalized seeing size (FWHM).
In this study, $I_{th}$ corresponds to $24.23$ mag/arcsec$^2$ and
$N_{min}$ to $30$ pixels.
We detect objects and measure several photometric parameters simultaneously.
For example, the important parameters we measure are; cx, cy: image centroids (in units of pixels), $f_{iso}$: total flux above the threshold isophotal surface brightness, $\npix$: number of connected pixels, $\peak$: counts at the peak brightness, $SB$: mean surface brightness, $q$: axis ratio, $p$: position angle, $\cin$: central concentration index, and flag for blended objects.
$\cin$ is defined as

\beq
\cin\equiv{\int\limits_0^{\alpha r(\mu_L)} rI(r)dr}/
          {\int\limits_0^{       r(\mu_L)} rI(r)dr},
\eeq

\noindent
where $r(\mu_L)$ is the ^^ equivalent radius'\footnote{Abraham et
al. (1994) proposed a modified scheme using elliptical apertures
based on the second moment of images, which is in fact
practically identical to our scheme described by Doi et al. (1993).} at a limiting surface brightness $\mu_L$, and $\alpha$ is a constant ($0< \alpha <1$).
We take $\alpha=0.3$ following Doi et al. (1995).
Both the axis ratio $q$ and the position angle $p$ are measured on the
basis of the profile within the isophote at $I_{th}$.
An object is identified as blended, if it has multiple peaks associated with more than $\npix/10$ pixels when it is sliced step by step toward higher isophotal threshold levels.
Objects that are flagged as blended are not included in the sample in this study.
The parameter $f_{iso}$ gives the isophotal magnitude (section $3.7$), $\peak$ and $\npix$ are used in the star/galaxy discrimination (section $4.1$), and $\cin$ and $SB$ are used in morphological type classification (section $4.3$).     
\subsection{Photometric Calibration}

Photometric zero-point calibration is carried out as an off-line process using photometry of standard stars.
We observed standard stars in the globular cluster NGC6205 (\cite{fd86}).
Its size was smaller than the field of view of one CCD chip.
Therefore in this calibration, we are concerned with the data taken by
a single chip with which we observed standard stars.
Since we have a consistent flux scale all over the contiguous field by mosaicking, the zero point of the chip applies to all the field.
Proper correction for atmospheric extinction requires the color of galaxies.
However, since we have $R$-band data only, we assume ($V-R$)$=0.77$ for all the galaxies we detected.
The value $0.77$ is the weighted mean of the colors of local galaxies of
various morphological types (\cite{yt88}).
The lack of a color term would make a relatively large photometric uncertainty.
We adopt magnitude correction $A^b_R=0.03$ for Galactic absorption according to Burstein \& Heiles (1984), which is based on the HI column density.

\subsection{Astrometric Calibration}

The Guide Star Catalog (\cite{las90}, hereafter GSC) is used as a reference to transform positions of objects given by the internal coordinates to the celestial coordinates.
GSC contains the positions (J$2000.0$ epoch) of $1.7\times10^7$ stars all over the sky and the accuracy is claimed to be in the range $0.^{\prime\prime}2$ to $0.^{\prime\prime}8$ all over the catalog.
Common stars cataloged both in our sample and in GSC are found by cross-matching.
We have $355$ common stars which are almost uniformly distributed in our field.
The saturated stars are excluded.
The stars we use for reference have errors $0.^{\prime\prime}3$ to $0.^{\prime\prime}5$, which are sufficiently accurate for our purposes.
The standard coordinates ($\xi$, $\eta$) of the reference stars are
expressed by the $2$nd-order polynomials of the measured (x, y)
coordinates, and the plate constants were determined by the least
squares method.  
In the fitting procedure, stars with large residuals are rejected.
We use $260$ stars for the final fitting.
Then ($\xi$, $\eta$) is transformed to ($\alpha$, $\delta$).
After the transformation, no systematic trend is seen in the residual vectors, except that large ($<3.^{\prime\prime}5$) residuals are seen near the southeast corner.
Since the data are lacking in the southwest corner, the mosaicking
procedure of southeast frames is supposed to be less accurate compared with that of other frames.
The residual of the fit to the reference stars lies in the range
$-2.^{\prime \prime}5$ -- $3.^{\prime \prime}5$ (r.m.s. error is $\sim 1^{\prime \prime}$).
The errors are small enough and do not affect the measurement of the angular correlation function.

\section{SAMPLE SELECTION}

\subsection{Star/Galaxy Discrimination}

The catalog of detected objects contains stars, galaxies, cosmic rays,
dusts or defects on the CCD chips, and other noises.
Our technique for distinguishing galaxies from stars is based on the difference in image profile that the image of a galaxy is more extended than that of a star.
The stellar image is determined by the instrumental optics and the seeing, and therefore lies in a well-defined region of an appropriate parameter space.

We take a parameter log($\peak$/$\npix$) as a star/galaxy discriminator where $\peak$ is the peak count of an object image and $\npix$ is the number of connected pixels above a given threshold.
Stars are expected to have a higher log($\peak$/$\npix$) than galaxies at the same magnitude.
The diagram we use for star/galaxy discrimination is shown in Figure~\ref{fsgsep}.
Stars form a well-defined sequence running from top-left to middle-right in Figure~\ref{fsgsep}, while galaxies are distributed extensively below the star sequence indicating that galaxy profiles have more significant variations.
The star sequence merges into the galaxy locus at $m_R \sim 18.5$.
Faint stars become indistinguishable from galaxies because of seeing and increasing noise.  
The objects which are distributed above the merged star/galaxy sequence in the faint region ($m_R \geq 19.0$) were confirmed to be defects, noises, or cosmic rays.
The star sequence terminates rather abruptly at the bright end ($m_R < 15$).
This is due to the fact that we excluded the bright saturated stars. 
Since some of the brightest galaxies are also excluded due to saturation, our catalog is not complete for the brightest galaxies ($m_R \leq 13.5$).

We define the boundary line between stars and galaxies, which is drawn manually and denoted by the solid line in Figure~\ref{fsgsep}.
Then we compare the star counts with those of Yamagata \& Yoshii (1992, hereafter YY92) model for the same galactic coordinates as the Coma cluster region.
This model is more sophisticated than the model given by BS83.
The YY92 model is based on number counts of stars with more precise
and deeper observations, and the thick disk component is introduced
in the model (Yoshii 1994, private communication).
In Figure~\ref{fystar}, the observed star counts (filled circle) based
on the boundary we define show very good agreement with the YY92 model
(solid line), while observed star counts disagree with the BS83 model (dotted line).
The faint limit of our galaxy sample extracted with this star/galaxy discrimination method is determined to the magnitude at which our star counts becomes inconsistent with the YY92 model.
The faint limit is $m_R\sim18.5$ . 
This limiting magnitude is largely influenced by the seeing size. 

\placefigure{fsgsep}
\placefigure{fystar}

\subsection{Estimate of the Number of Field Galaxies}

In order to investigate the properties of cluster galaxies, it is necessary to estimate the contamination by background and foreground galaxies, which we call ^^ field' galaxies hereafter.
We estimate the number of contaminating field galaxies with our own
imaging data for SA57. 
The center position of the field we surveyed is ($13^h 09^m 46.^s6$, $+29^{\circ} 23^{\prime} 02^{\prime\prime}$) (J$2000.0$), which is $2.^{\circ}65$ away from the Coma center.
The data were taken by our second mosaic CCD camera, MCCD2 (Kashikawa
et al. 1995a; Okamura et al. 1997) at the prime focus of the William
Herschel Telescope in $1996$ Apr.
MCCD2 is basically the same instrument as MCCD1.  
The total survey area was $0.44$ square degree with $1.^{\prime\prime}0$ seeing which
is much better than those of the Coma data.
The SA57 data were smoothed with a Gaussian beam so as to have the
seeing size $\sim 4.^{\prime \prime}5$ of the Coma data.
We apply to the SA57 data the same manner of data reduction, the same
detection threshold, and the same star/galaxy-separation method as
those for the Coma data.

The expected contamination rate, \ie the number of field galaxies with
respect to that of our Coma galaxies is listed in Table \ref{tbgc} as
a function of the limiting absolute magnitude.
We restrict our sample to galaxies brighter than $m_R^{lim}=18.21$
($M_R^{lim}=-16.00+5$log$h$), where the contamination rate reaches 
$\sim50\%$.
We here assume the distance modulus of the Coma cluster to be
$(m-M)_0=34.21$ which is derived from the mean velocity $\overline
v=6942$kms$^{-1}$ (\cite{zgh93}).
Throughout this paper we adopt a Hubble constant $H_0=100h$km/s/Mpc.
We hereafter use $h=1$ where the $h$ dependence is not explicitly indicated. 

\placetable{tbgc}

\subsection{Morphological Classification}

We classify the sample galaxies into two subsamples using
a single consistent parameter, the degree of luminosity 
concentration, to parameterize the morphology of galaxies over
the wide magnitude range $-20 \leq M_R \leq -16$, where both giant and 
dwarf galaxies are included. 
One consists of galaxies which have high central concentration of
surface brightness
(hereafter HCC), and the other consists of those with low central 
concentration (hereafter LCC). 
As for the giant ($M_R\leq-18$) galaxies, the former generally corresponds to early-type
galaxies (Hubble morphological type Elliptical/S0, the type index $T=-6$ -- $-1$) and the latter to late-type galaxies (Spiral, $T=1$ -- $10$).
This method performs a crude classification using the known properties
of giant galaxies that early-type galaxies show de Vaucouleurs' surface-brightness profile which has higher central concentration (corresponds to larger $\cin$ in our measured parameters, see section $3.2$) than the exponential profile of late-type galaxies.  
The basic idea of this method is described in Okamura, Kodaira, \& Watanabe (1984) and extended by Doi, Fukugita, \& Okamura (1993) who used the parameters $\cin$ and $SB$ to classify galaxy morphology.
They showed that the method is robust for a variety of image size and inclination of galaxies.
Note that both of these parameters are distance independent.

As for the morphological classification of dwarf galaxies ($M_R>-18$),
on the other hand, there is no generally accepted quantitative
definition, which sometimes leads to confusion (e.g., \cite{fb94}).  
In this study, we classify both giant and dwarf galaxies on the basis 
of the single objective photometric parameter $\cin$.  
However, we extrapolate the critical line between the bright HCC and 
the bright LCC galaxies to dwarf galaxies as described below.  
Readers should be careful that the HCC/LCC dwarfs do not correspond to 
so-called ^^ early-type dwarfs'/ ^^ late-type dwarfs'.  
It might be more relevant to suppose that the
faint HCC corresponds to the ^^ nucleated dE', and the faint LCC to
the ^^ non-nucleated dE', ^^ dS0', and ^^ dwarf spheroidals'.
However, it is unclear at the moment what our HCC/LCC dwarfs actually
mean in terms of these morphological types which are based on the eye
inspection.
A similar analysis of, for example, the Virgo dwarfs using
high-resolution imaging data would be useful to understand this.

In practice, we compute $\cin$ for two series of model galaxies with
different magnitudes, surface brightness, and axis ratios; one is for
galaxies with de Vaucouleurs' law profile ($\cin^E$) and the other for
galaxies with the exponential law ($\cin^S$).
The parameter ranges are taken so that they cover those of our sample galaxies, \ie $14.0\leq m_R \leq 20.0$, $ 21.0\leq SB \leq 24.0$, and $0.2 \leq$ axis ratio $\leq 1.0$.
The model galaxies are smeared with the seeing profile (a single Gaussian) with our observed seeing size ($\sim 4.^{\prime\prime}5$).
We obtain the model-predicted $\cin$ for de Vaucouleurs' law profile and the
exponential law profile for each sample galaxy by ^^ measuring' the
model profile with the same threshold as for the real sample.

In the Coma cluster sample, we identify some bright galaxies whose morphological types are given in Dressler (1980a). 
These bright galaxies are shown in Figure~\ref{fessep} in the
$\cin$ versus magnitude space as well as the locus of the two series
of the models.
It is found that the model prediction is consistent with Dressler's classification.

With the model galaxies, we define for each sample galaxy the critical
value $\cin^{crit}$ by which we classify it either into the HCC type
or into the LCC type as 

\beq
\cin^{crit} = \cin^S + p \times (\cin^E - \cin^S),
\eeq

\noindent
where $p$ is a free parameter, and $\cin^E$ and $\cin^S$ are the values of $\cin$ of model galaxies with de Vaucouleurs' law profile and the exponential profile, respectively, that have the same magnitude, same surface brightness, same seeing size, and same axis ratio as the sample galaxy.  
We classify a galaxy as the HCC/LCC type when it has $\cin$ larger/smaller than $\cin^{crit}$. 
We determine the parameter $p$ empirically as follows using the Dressler's classification as fiducial.
We investigate the completeness of the subsamples by changing $p$ from $0.0$ to $1.0$.
Here we define the completeness of the sample $C_{e}(p)$ for early
types and $C_{s}(p)$ for late types as
 
\beq
C_{e}(p) = \frac{N_{ee}}{N_{ee}+N_{se}}, \hspace{3mm}{\rm and}\hspace{3mm} 
C_{s}(p) = \frac{N_{ss}}{N_{ss}+N_{es}},
\eeq

\noindent
respectively, where $N_{ee}$ is the number of Dressler's early-type
galaxies classified correctly as the HCC type at given $p$, $N_{se}$
Dressler's late-type galaxies mis-classified as the HCC type, $N_{ss}$
Dressler's late-type galaxies classified correctly as the LCC type, and
$N_{es}$ of Dressler's early-type galaxies mis-classified as the LCC type.
Both $C_{e}(p)$ and $C_{s}(p)$ are shown in Figure~\ref{fcomp} as a function of $p$.
The completeness of early-type sample $C_{e}(p)$ decreases with $p$,
while $C_{s}(p)$ increases with $p$.
Both $C_{e}$ and $C_{s}$ change smoothly with $p$ and are well determined regardless of the magnitude.
To obtain an impartial sample for both early-type and late-type
galaxies, we choose $p$ that makes $C_{e}(p)$ equal to $C_{s}(p)$
as the ^^ standard' value.
The standard $p$ is chosen to be $0.35$ (Figure~\ref{fcomp}).
We call this sample with $p=0.35$ as the ^^ standard sample'.
Figure~\ref{fes} shows the magnitude versus $\cin$ plot for the standard sample.
Galaxies classified as the HCC type are indicated by filled circles
and those as the LCC type by open circles.
  
To evaluate the uncertainty in this morphological classification, we
make two additional samples; one is the ^^ upper sample' which is
made with $p=0.40$, and the other is the ^^ lower sample' with $p=0.30$.
In the upper sample, we can obtain a higher purity in the sample of
the HCC-type galaxies at the cost of lower completeness, while in the
lower sample, we obtain a higher purity in the sample of the LCC-type galaxies.
The influences of this uncertainty will be discussed in section $6.2$.
\vs

\placefigure{fessep}
\placefigure{fcomp}
\placefigure{fes}

The sky distribution of galaxies is shown in Figure~\ref{fsky} for the
total, HCC, and LCC samples.
A larger symbol denotes a brighter galaxy.
It is evident that the bright HCC-type galaxies have a more centrally concentrated distribution on the sky which is consistent with the morphology-density relation (\cite{dre80b}).
This suggests that our classification works well.

\placefigure{fsky}

\vs
Among the three limiting magnitudes, \ie the limit of star/galaxy
discrimination ($M_R=-15.7$), that due to the contamination of the
field galaxies ($M_R=-16.0$), and that of morphological classification
($M_R=-15.5$), the brightest limit ($M_R=-16.0$) should determine the limiting magnitude of the final sample for our analysis.
The total number of galaxies cataloged is $1822$, and morphological
contents are $1313$ for the HCC and $509$ for the LCC.
The luminosity function is given in Kashikawa et al. ($1995b$) together with those of three other nearby clusters. 

\section{RESULTS}

\subsection{Luminosity Segregation}

\subsubsection{Angular Correlation Function}
To quantify the galaxy clustering, we use the angular two-point
correlation function $\acf$, which is usually used for estimating the
clustering of field galaxies. 
Most of previous studies used radial profiles to investigate the 
clustering properties in rich clusters assuming the symmetric
structure of clusters. 
However, quite a few clusters are known to show
asymmetry or substructures and the Coma cluster is no exception
(e.g., \cite{bhb92}; \cite{cal93}).
The $\acf$, which quantifies any sort of distributions for different 
correlation scales, is hardly affected by asymmetry and therefore considered
to be a robust measure of clustering properties in clusters as well
as fields.

The $\acf$ is defined by $\delta P_{12}$,
the differential probability of finding a pair of galaxies, one in solid angle
$\delta\Omega_1$, and the other in $\delta\Omega_2$ separated by $\theta_{12}$:

\beq
\delta P_{12} = n^2[1+\omega(\theta_{12})] \delta\Omega_1 \delta\Omega_2,
\eeq 

\noindent
where $n$ is the average number density of galaxies (\cite{pee80}).
In practice, $\acf$ is derived by,

\beq
\acf = \frac{N_{gg}(\theta)}{N_{gr}(\theta)}\frac{2N_r}{(N_g-1)}-1
, 
\eeq 

\noindent
where $N_{gg}(\theta)$ is the number of pairs of galaxies separated by
$\theta$, $N_{gr}(\theta)$ is the number of pairs of a galaxy and a random point separated by $\theta$ as well, $N_g$ is the number of galaxies, $N_r$ is the number of random points.
The random points are generated so that they are distributed in the area identical in shape to the area we surveyed.
Circular regions around several bright stars, and rectangular regions where pixel defects are present are excluded from the area of the random sample, in the same manner as in the real sample.
In each random sample, random points are generated ten times the number of sample galaxies, and $N_{gr}(\theta)$ evaluation is made for twenty different random samples.  
Therefore, errors caused by the random number generation are negligible.
 
Our code is first checked with random samples to confirm that $\acf
\sim 0$ over all scales, and then checked with the CfA sample (\cite{lgh88}) to make sure that our code reproduces the same $\acf$ as they derived.

The $\acf$ of field galaxies is well approximated by a power law

\beq
\acf = A_{\omega}\theta^{\gamma}.
\eeq

\noindent
Though it is not sure that this approximation is also adequate for
cluster galaxies, we here take the same power law fitting.  
We can quantify the clustering properties by two parameters, the amplitude $A_{\omega}$ and the power index $\gamma$.
In addition to $A_{\omega}$ and $\gamma$, we introduce a single
parameter $W_{int}$ as the integration of $\acf$ from $\theta_1$ to
$\theta_2$:  

\beq
W_{int}=\int\limits_{\theta_1}^{\theta_2} \acf d\theta.
\eeq

\noindent
We choose ($\theta_1$, $\theta_2$) as ($0.^{\circ}1$, $1.^{\circ}0$)
corresponding to the physical scale ($0.12h^{-1}$Mpc, $1.20h^{-1}$Mpc) 
at the Coma distance.
 
The sampling error in two point correlation function is usually larger than
the error due to Poissonian fluctuation (\cite{mjb92}).  
The $1\sigma$ errors in these parameters $A_{\omega}$, $\gamma$, and $W_{int}$, and error bars of each bin are estimated by the bootstrap resampling method (hereafter BRM; \cite{de83}; \cite{bbs84}). 

We extract three subsamples of different absolute magnitudes from the
total sample, the HCC sample, and the LCC sample: $M_R \leq -18.0$
(the bright sample), $-18.0< M_R \leq -17.0$ (the intermediate
sample), and $-17.0 < M_R \leq -16.0$ (the faint sample).
Since some of the brightest galaxies ($M_R \leq -20.7$) are saturated, 
we excluded them from the bright sample. 
    
\subsubsection{Luminosity Segregation in the Total Sample}

Figure~\ref{facfls}(a) represents the $\acf$s of the
three subsamples extracted from the total sample.
The $\acf$s are computed for separations up to the scale at which the amplitude of $\acf$ falls to negative.
The solid line, the dashed line, and the dotted line indicate the
$\acf$ for bright sample, intermediate sample, and faint sample, respectively.

Figure~\ref{facfls}(a) shows an important result that the amplitude of the $\acf$ varies with luminosity. 
This demonstrates that the degree of clustering shows a pronounced
dependence on luminosity of galaxies: brighter galaxies have a much larger correlation amplitude than fainter ones.
The parameters, $\gamma$, $A_{\omega}$, $W_{int}$, and $\delta W_{int}$, \ie the $1\sigma$ error in $W_{int}$ estimated by BRM, are listed in Table \ref{tacfls}.
A systematic trend is found in $A_{\omega}$ and $W_{int}$ that a
brighter sample tends to have a larger $A_{\omega}$ and a larger $W_{int}$. 
Therefore, the LS is confirmed for the total sample.

The $\acf$s of all the subsamples, especially the bright sample, are
flattened at smaller scales ($\theta \leq 0.^{\circ}1$), which is
consistent with the result obtained by Davis \& Geller (1976).
They showed that the slope of the $\acf$ becomes steeper when galaxies in dense cluster regions are excluded.
Brainerd \& Villumsen (1992) calculated the spatial two-point correlation function based on $N$-body simulations of the CDM model and showed that the function becomes flat in high density regions such as clusters of galaxies.
The leveling off of the correlation function in the dense cluster
region would suggest that clustering properties in rich clusters would be influenced more or less by some non-linear effect of mutual interactions of galaxies or global dynamics in clusters.

\placefigure{facfls}
\placetable{tacfls}

\subsubsection{Luminosity Segregation in the Morphological Subsamples}

We show in Figure~\ref{facfls}(b) and in Figure~\ref{facfls}(c) the
$\acf$s of the three subsamples extracted from the HCC sample and the
LCC sample, respectively.  For the HCC galaxies, the bright sample has
a stronger degree of clustering than the faint sample.  The tendency
is similar to that seen in the total sample discussed above.  The
difference in the clustering amplitude between the bright sample and
the faint sample is, however, larger than that found for the total
sample. On the other hand, such a tendency can be barely seen for the LCC
galaxies (Figure~\ref{facfls}[c]).  These characteristics are
quantitatively seen in the $\acf$ parameters listed in Table
\ref{tacfls}.

Thus, we conclude that the LS is strong in the HCC-type galaxies and
weak in the LCC-type galaxies in the Coma cluster.
The reason why the LS is observed for the total sample is that $\sim 3/4$ of
the total sample is comprised of the HCC type which shows strong LS.
In terms of the morphological segregation, Figure~\ref{facfls} and
Table \ref{tacfls} clearly demonstrate that galaxies of different
morphological types have different clustering properties even in the
same range of luminosity.

In addition to the clustering behavior of bright galaxies, we can
investigate that of {\it faint} galaxies ($M_R>-18$) based on the
intermediate sample and faint sample. This is a unique feature of the
present study. In the intermediate sample, there is no significant
difference in the correlation amplitude between HCC and LCC galaxies.
On the other hand, HCC galaxies in the faint sample show a slightly
weaker correlation amplitude than LCC galaxies in the same sample,
though the difference is statistically marginal.
Thus, dwarf galaxies do not show as strong morphology
segregation as giant galaxies, if the morphology of dwarf galaxies is
characterized by the degree of luminosity concentration.
       
In order to see more clearly the dependence of the clustering strength
on limiting magnitude $M_R^{lim}$, we plot $W_{int}$ as a function of
$M_R^{lim}$ in Figure~\ref{fwintms}.  The solid line denotes the
HCC-type galaxies and the dotted line denotes the LCC-type galaxies.
The error bars show the fluctuations estimated by BRM.  The HCC-type
galaxies show a strong dependence on the limiting magnitude such that
a sample with a brighter limiting magnitude has a larger $W_{int}$.
On the other hand, the $W_{int}$ of the LCC-type galaxies has a weak
dependence on the limiting magnitude.

As far as the {\it bright sample} is concerned, the above findings are
qualitatively accounted for by the well-known morphological segregation. 
The bright HCC-type galaxies are mostly E and S0 galaxies and the bright
LCC-type galaxies are mostly spirals. E galaxies stretch to higher luminosity than S0 galaxies. And, E galaxies most favor high density regions, and 
spiral galaxies least with S0 galaxies in between. These properties
explain the larger correlation amplitude of the HCC-type galaxies.
The increase in the correlation amplitude of the HCC sample with
increasing luminosity can also be accounted for by a larger
fraction of E galaxies over S0 galaxies.
However, such explanations based on the known properties of bright
galaxies are not relevant to the behaviors of the correlation amplitude 
of the {\it intermediate and faint samples}.

\placefigure{fwintms}

\subsection{Dependence on the Clustercentric Radius}

In this section, we examine the properties of galaxies in terms of the cluster structure.
Figure~\ref{fradp} shows the surface number density (per square
degree) as a function of the radius from the cluster center.
Our survey field has a rectangular shape and it includes some rejected regions with complicated boundaries.
We estimate the net area indirectly by counting the artificial random points which are scattered over the effective survey field.
We scatter $10^5$ random points.
We confirm that the counted number of the points for each bin is
stable for this random number generation. 
The cluster center is defined as the position of the midpoint between
the two cD galaxies.

The surface number densities of all the subsamples increase toward the cluster center.
The bright HCC-type galaxies are strongly concentrated on the cluster
center, while the fainter HCC-type galaxies have flatter density
profiles.
The density profiles of the LCC-type galaxies weakly depend on luminosity.
In the faint subsamples, the density profile of the HCC-type galaxies is
similar to that of the LCC-type galaxies.
If our faint HCC/LCC galaxies correspond to nucleated/non-nucleated
dwarfs, this result is different from that found for dEs in Virgo and
Fornax clusters (\cite{van86}, \cite{bts87}, \cite{fs89})  
It is also noted that their samples are restricted to small scales,
$\leq 1.4h^{-1}$Mpc, while ours extends to $\sim 3.0h^{-1}$Mpc.
Quintana ($1979$) showed that the core radius of the Coma cluster changes significantly with the limiting magnitude of the sample: larger core radii for deeper samples.
His finding is consistent with our result.
We note, however, that the magnitude dependence of core radius is
stronger for the HCC-type galaxies than for the LCC-type galaxies.

\placefigure{fradp}

\section{RELIABILITY OF THE RESULT}

\subsection{Effects of Contamination by Field Galaxies}

As mentioned in section $4.2$, our samples are contaminated by non-cluster members, most of which are supposed to be galaxies behind the cluster.
The amplitude of $\acf$ of a faint sample which contains a larger
fraction of field galaxies should be smaller than that of a bright
sample including a higher fraction of cluster members because of the following two reasons. 
One is that field galaxies have an intrinsically weaker clustering amplitude than cluster members.
The other is that the amplitude of the $\acf$ depends on the effective
depth of the sample in such a way that a deeper sample has a weaker
amplitude in small scales than a shallower sample, because the same physical length subtends a smaller angle at larger distances, and that more galaxies are seen in projection (\cite{gp77}).
The expected contamination rates of field galaxies of our samples are
up to $50\%$ (see section $4.2$).
There remains a possibility that we have mis-classified many field
galaxies as HCC types.
We undertake the following test to evaluate statistically the effects of field galaxies on our results.

We test whether or not the amplitude difference in $\acf$ we observe
is caused only by the contamination of field galaxies.
As described above, the amplitude of $\acf$ depends on the sample depth.
The dependence is known to be expressed by a scaling relation of the relativistic version of Limber's equation (\cite{gp77}).
It is necessary to include into the scaling relation not only redshift effect and curvature effect but also the effect of luminosity evolution of galaxies when we are concerned with a deep galaxy sample (e.g., \cite{sha80}).
In so-called ^^ scaling test', we investigate whether or not the $\acf$s of two samples with different sample depths coincide with each other after one of them is scaled to the other on the basis of the scaling relation.
We apply the scaling test to our samples on the assumption that our sample includes some fraction of field galaxies whose sample depth is  deeper than that of cluster members.

Consider a sample of galaxies which is composed of cluster members
(fraction $f_c$) and of field galaxies (fraction $f_f=1-f_c$).
Suppose that cluster members and field galaxies have their intrinsic
angular two point correlation functions $\omega_{cc}(\theta)$ and
$\omega_{ff}(\theta)$, respectively.
The observed $\acf$ of the sample is expressed as

\beq
\omega(\theta) = f_{c}^2 \omega_{cc}(\theta) + f_{f}^2 \omega_{ff}
(\theta) + 2f_{f}f_{c} \omega_{cf}(\theta). 
\eeq 

\noindent
We assume no cross-correlation between cluster members and field
galaxies, i.e., $\omega_{cf}(\theta)= 0$.
We assume that $\omega_{cc}$ is independent of the sample depth, \ie
the limiting magnitude of member galaxies in the sample.
Therefore $\omega_{cc}$ is not subjected to scaling and only $\omega_{ff}$ is scaled according to the sample depth.
Previous estimates of the $\acf$ of field galaxies show that $\omega_{ff}$ is approximated by a simple power law as

\beq
\omega_{ff}(\theta)=A\theta^{-0.8}.
\eeq  

\noindent
The sample depth can be estimated by the galaxy surface number density of the sample (\cite{gp77}).
The scaling relation between the amplitude $A$, which is the amplitude of $\omega_{ff}$ at $\theta=1^{\circ}$, and the galaxy surface number density, which is a measure of the effective sample depth, was derived by Shanks et al. (1980) for a sample of galaxies in 
the $R$ band.
Their sample, however, covers a deeper magnitude range than our samples and we extrapolate their relation to our brighter magnitude range.
The luminosity evolution of galaxies is taken into account in their scaling relation, but it gives no difference in the magnitude range we are concerned.

We construct three new subsamples with different limiting magnitudes, $M_R^{lim}=-18.00$ ($16.21$ in apparent magnitude), $-17.00$ ($17.21$), and $-16.00$ ($18.21$), respectively.
We estimate the surface number density of galaxies (per square
degree), $N_{exp}$, at the limiting magnitudes of the respective
samples with our galaxy counts in the SA57 explained in section $4.2$.
Then, we derive the predicted amplitude $A$ of each subsample using the extrapolated version of the scaling relation by Shanks et al. (1980).
The fraction of field galaxies in each subsample ($f_{f}$) is
calculated on the basis of our estimate of the number of field galaxies described in section $4.2$.
The predicted value of $A$ is given in Table~\ref{tst} together with $N_{exp}$, $f_{f}$, $f_{f}^2A$, and $f_c^2\omega_{cc}$ for each subsample.
Note that the contributions by field galaxies ($f_{f}^2A$) to the
amplitude are almost the same for the three subsamples and are negligible ($f_f^2A<10^{-2}$) compared with the observed amplitude at $\theta=1^{\circ}$ ($\omega > 10^{-2}$, see Figure~\ref{facfls}).
Thus, we scale the observed $\acf$ according to the values of $f_c^2\omega_{cc}$ neglecting the term $f_f^2\omega_{ff}$ in equation ($8$). 
If the difference in amplitude among the subsamples with different limiting magnitudes is totally due to the contamination of field galaxies, the scaled $\acf$ of the subsamples would agree with each other.  

Figure~\ref{fstest} shows the $\acf$s of the three subsamples with different limiting magnitudes.
The observed $\acf$ of the subsample with a brighter limiting magnitude
(open circles) is scaled so that the scaled $\acf$ (filled circles)
coincides with the observed $\acf$ of the subsample with a fainter limiting magnitude if the difference between the two $\acf$s is totally due to the contamination of field galaxies.
The $\acf$s of two subsamples with brighter limiting magnitudes are scaled to that of the subsample with the faintest limiting magnitude.
The scaled $\acf$s still show a significant difference from the observed $\acf$ of the subsample with the faintest limiting magnitude.
This demonstrates that the observed strong LS is not due to the contamination of field galaxies.

\placetable{tst}
\placefigure{fstest}

\subsection{Uncertainty in Morphological Classification}

To evaluate the effect of the uncertainty in morphological classification,
we calculate the $\acf$s for the lower sample which has $p=0.3$ (see
section $4.3$) and the upper sample which has $p=0.4$.
Figure~\ref{ferr} shows $W_{int}$ as a function of $M_R^{lim}$ for the
lower sample (dotted line) and the upper sample (dashed line) as well
as the standard sample (solid line) with $1$ $\sigma$ fluctuations estimated by BRM.
The global behavior of $W_{int}$ for both the lower and the upper
samples is the same as that for the standard sample.  
The difference of $W_{int}$ between the standard sample and the lower/upper sample is almost the same as or less than $1$ $\sigma$
fluctuations of $W_{int}$ itself over the whole magnitude range
observed.
We conclude that the $\acf$ is insensitive to the errors in morphological classification.

\placefigure{ferr}

\subsection{Contamination by Stars}

Our faint samples are probably contaminated by stars.
We believe, however, that the fraction of mis-classified stars is
negligible because the limiting magnitude of our star/galaxy
separation reaches a much fainter magnitude than the limit imposed by
field galaxy contamination (section $4.2$), and the star counts are
almost consistent with the YY92 model (Figure~\ref{fystar}).
We have also confirmed that our star sample gives $\acf\sim0$ over all the scales. 

\section{DISCUSSION}

We find a strong dependence of luminosity segregation on morphology;
the HCC-type galaxies show strong LS while the LCC-type galaxies show
weak or no LS.

Our results are obtained for cluster galaxies distributed in an over
density region on scales of $\leq 1h^{-1}$ Mpc.
Almost all of the previous studies on the LS were based on cluster samples
and significant LS was found in most studies (e.g., \cite{qui79};
\cite{cap80}; \cite{dp88}; \cite{shg90}; \cite{gar92}).
On the other hand, among the previous studies based on the
spectroscopic data of field galaxies, some claimed the significance of
the LS (e.g., \cite{dav88}; \cite{ham88}; \cite{sd90}; \cite{ml91}; \cite{lov95}), 
while others cast doubt on the significance of the LS (e.g.,
\cite{ps87}; \cite{tul88}; \cite{ed89}).
In the context of our result, this disagreement in field galaxies might be due to the different morphological contents of the samples they used.

Deep photometric samples of galaxies without redshifts have been used
so far to investigate the evolution of galaxy clustering with $\acf$
(e.g., \cite{nwd91}; \cite{pi92}; \cite{vfd97}).
The change of amplitude of the angular two-point correlation function
also reflects the evolutionary history of galaxies (\cite{ks84}). 
In these investigations of angular two-point correlation, they assumed
that the fainter galaxies in the apparent magnitude are farther
from us and that the degree of galaxy clustering does not depend on
luminosity.
However in practice, the faint galaxies would consist of both distant
galaxies and nearby dwarf galaxies.  
Our results suggest that the LS, that is the dependence of clustering
properties on luminosity, should be taken into account when one
examines the evolution of galaxy clustering.
Our results also suggest that it is necessary to consider the type
(HCC/LCC) mix of sample in order to investigate the LS.
If the sample contains a large fraction of the HCC type, it would show
significant LS.
We find that the bright HCC (early type) galaxies in the Coma
cluster show the LS, while the bright LCC (late type) galaxies show no LS. 
If this is universal for clusters of galaxies which are elliptical
rich in general, a galaxy sample containing a high fraction of cluster
galaxies would show the LS.
Consequently, in the study of clustering evolution based on $\acf$, it is
important to take into account the fraction of cluster galaxies.  

\section{CONCLUSIONS}

We have carried out a wide-field galaxy survey in the Coma cluster
region with a mosaic CCD camera to study the clustering properties of
cluster members. 
We have investigated the luminosity segregation (LS) quantitatively by
measuring the angular two-point correlation function and radial
distribution over the magnitude range of $-20 \leq M_R \leq -16$,
where both giant and dwarf galaxies are included.
Our analysis of the galaxy distribution based on the
morphology-classified galaxy samples with the unprecedentedly deep
limiting magnitude has yielded the following main results:

\begin{enumerate}

\item We have found that the galaxies with a high central
concentration in surface brightness profile (the HCC type) have strong
luminosity segregation while the galaxies with a low central
concentration (the LCC type) show almost no luminosity segregation, \ie
the strength of clustering of the LCC-type galaxies does not depend on
luminosity.

\item We have found strong segregation in luminosity for the total
sample of Coma cluster galaxies.
This is because the majority of the total sample is comprised of
the HCC-type galaxies which show strong luminosity segregation.

\item Brighter HCC-type galaxies tend to more strongly concentrate near the
cluster center than fainter HCC-type galaxies, while the LCC-type
galaxies do not show such dependence on luminosity in the density profile.  

\end{enumerate}

We have shown that these results are tenable against the contamination
by field galaxies and uncertainty in our method of morphological
classification.  
Our results suggest that it would be necessary to consider this
type-dependence of the LS in the study of clustering evolution with $\acf$. 

In order to see the universality of our results for clusters of galaxies in general, further studies for other clusters are necessary to investigate the luminosity and morphology dependence of galaxy distribution.
It is also desired to assemble a large sample of spectroscopic data for clusters of galaxies to construct samples of confirmed cluster members.

Probably both of a primordial and an environmental effects would have
influenced the clustering properties of galaxies in clusters. 
It is desirable to theoretically evaluate the LS due to each effect quantitatively.
In addition, the observational investigation of the LS in high-z clusters would directly distinguish the effects, although morphological classification and field correction would become more difficult in such clusters.

\acknowledgments

We thank the staff of Kiso Observatory for their technical help and
assistance to the development of MCCD1 and observations.  We wish to
thank Y. Yoshii and T. Yamagata for calculating star counts based on
their model.  
We acknowledge the referee, A. Dressler, for valuable comments which
improved the paper significantly.
NK acknowledges the Fellowships of the Japan Society for
the Promotion of Science for Japanese Junior Scientists.  This work
was supported in part by Ministry of Education, Science, and Culture
through Grant in Aid (02554001, 04452016), by Toray Science
Foundation, National Astronomical Observatory, and Hayakawa Foundation
of the Astronomical Society of Japan.

\clearpage

\figcaption[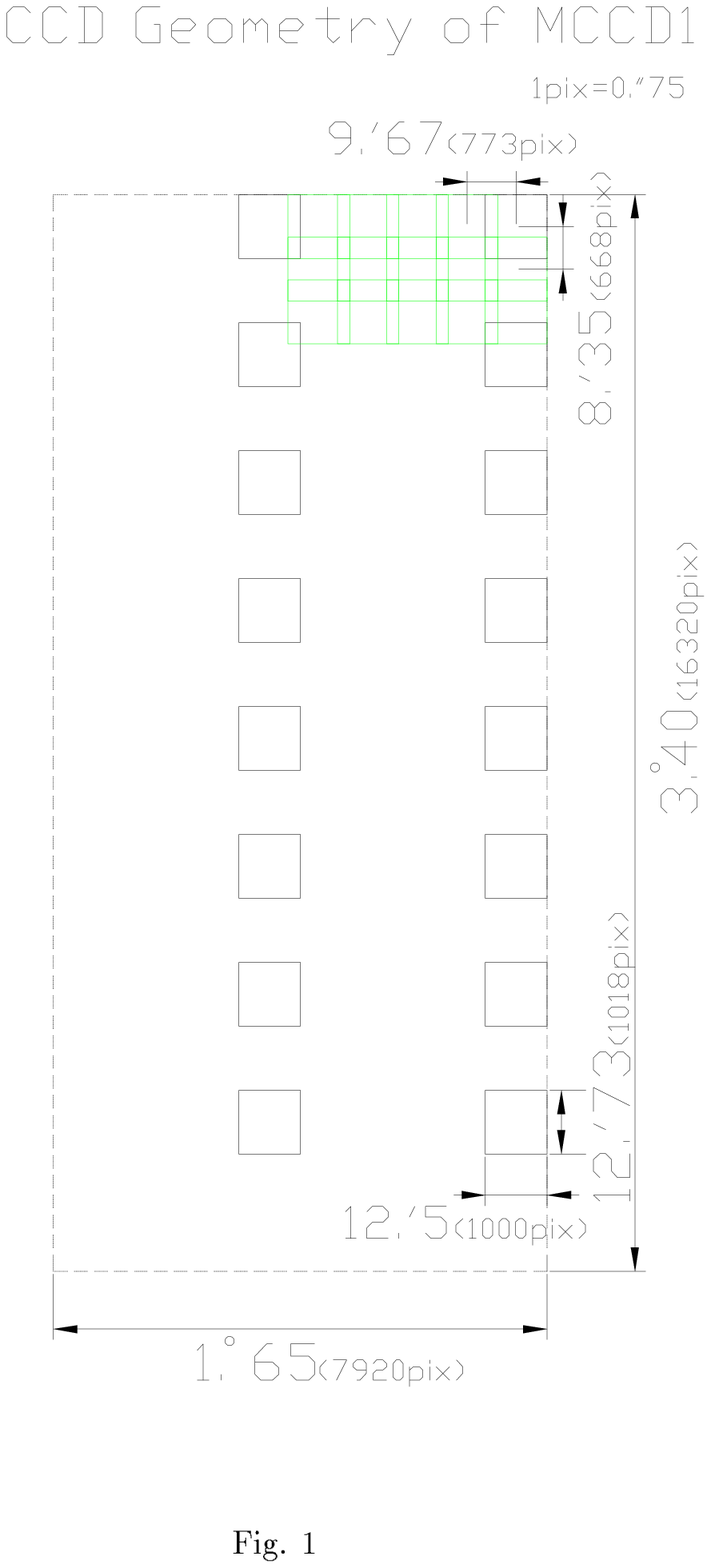]{Configulation of CCDs of MCCD1. 
Each CCD has $1000\times1018$ effective pixels.
The angular scale is shown when the camera is mounted on the prime focus of Kiso Schmidt.
North is up and west is to the left. 
\label{fcfg1}}

\figcaption[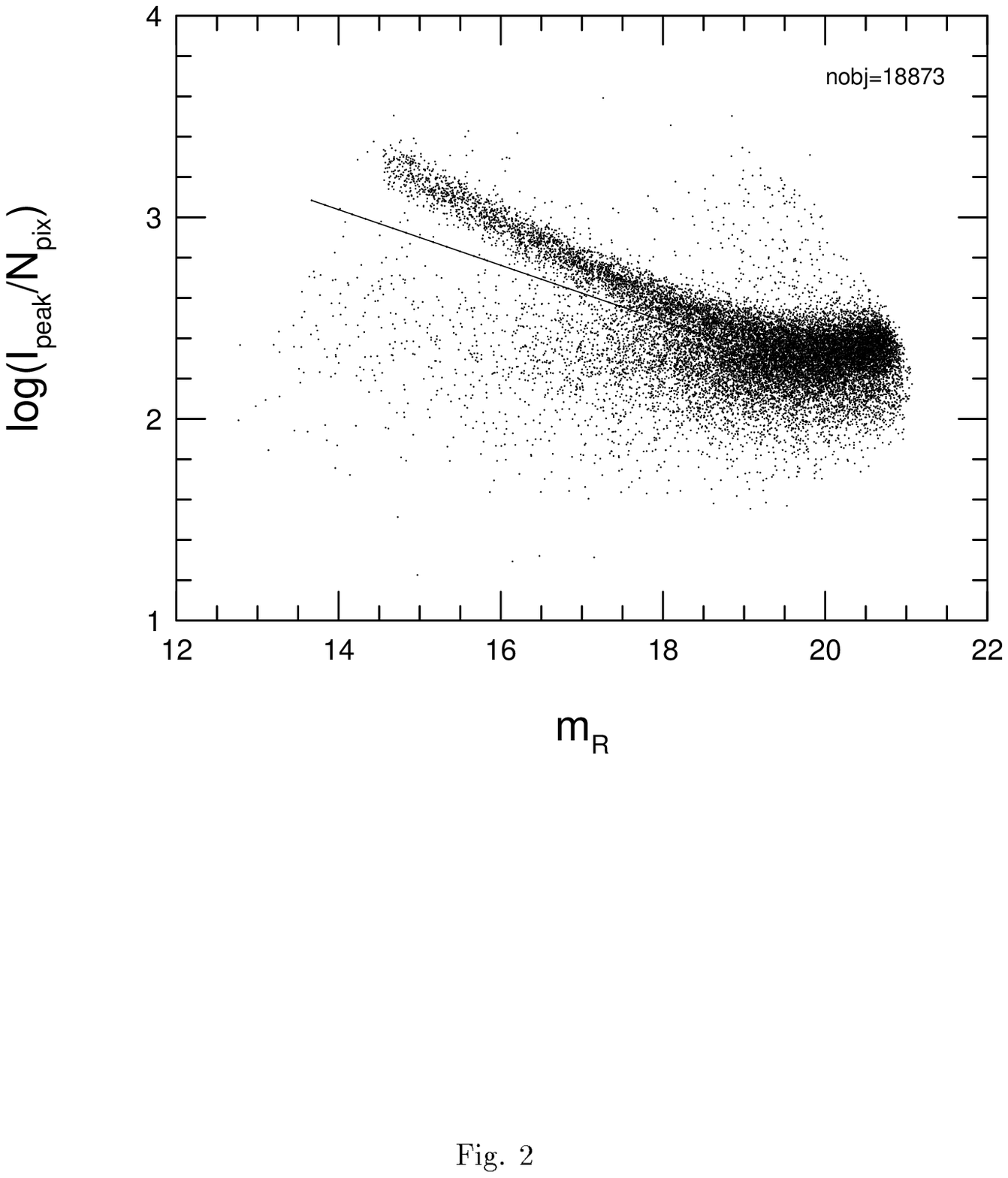]{Plot of log($\peak$/$\npix$)
vs. $R$-band apparent magnitude for the Coma cluster field. Stars form a well-defined continuous sequence running from top-left to middle-right in the panel.
Galaxies are distributed widely below the star sequence.
The solid line is the boundary we adopt for star/galaxy discrimination.
\label{fsgsep}}

\figcaption[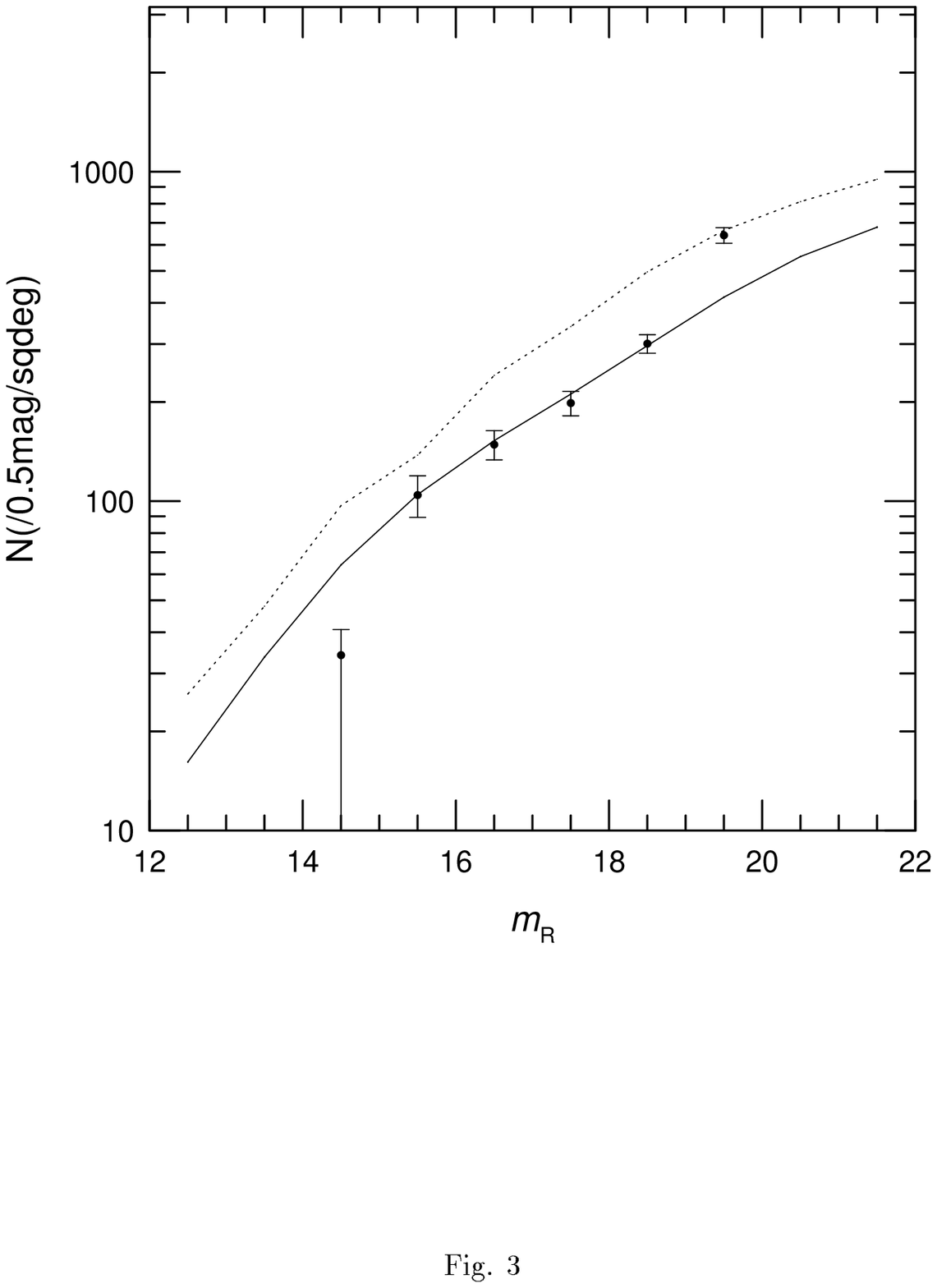]{Comparison of our star counts with
YY92 model and BS83 model.
Filled circles indicates the observed star counts based on the boundary line
we define. 
The solid line is for the prediction of the model by
Yamagata \& Yoshii (1992), and the dotted line is for that by Bahcall \& Soneira (1983).
\label{fystar}}

\figcaption[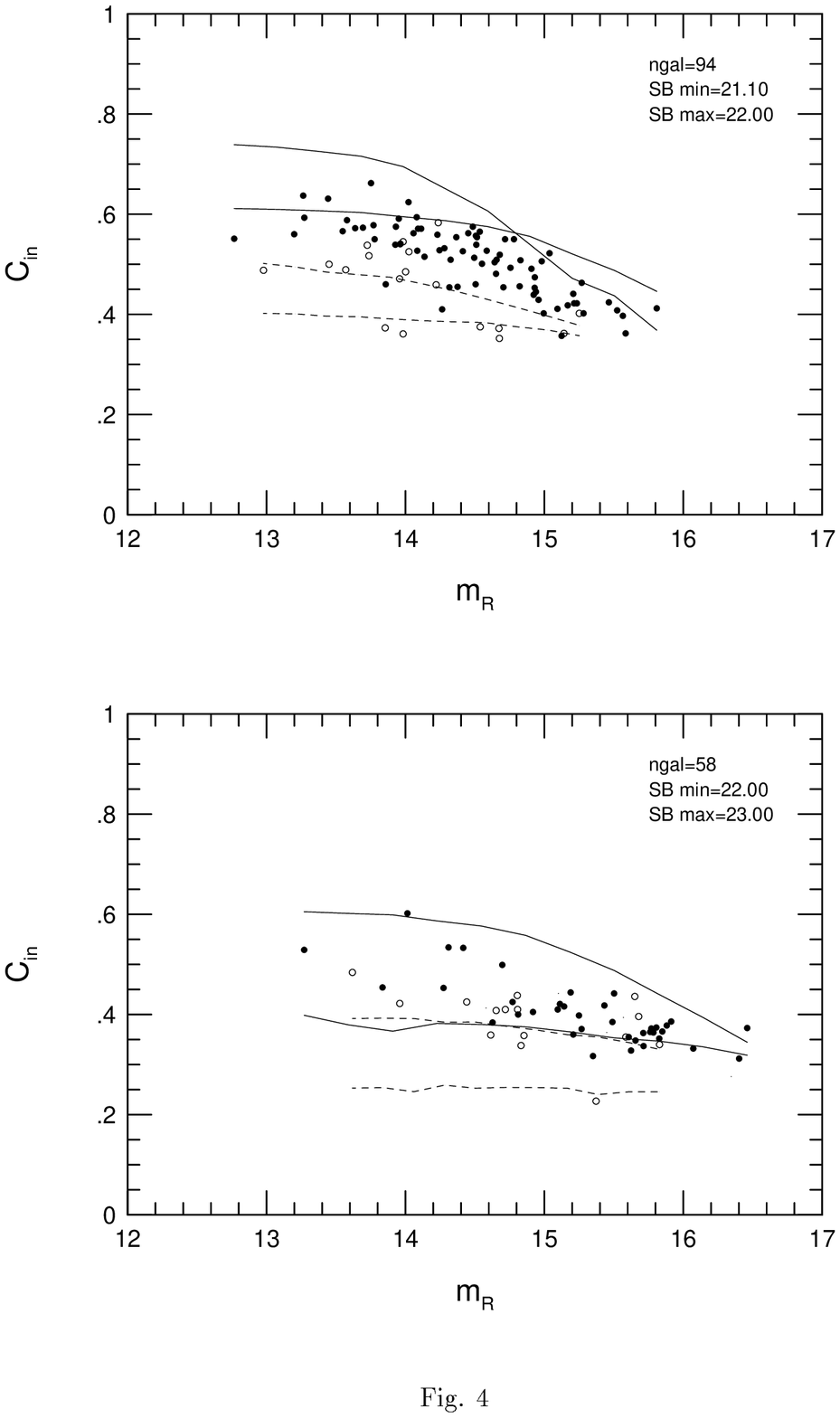]{$m_R-\cin$ plot of Dressler's sample.
The pair of the solid lines and the pair of the dotted lines
correspond to the loci of the HCC-type model galaxies and the LCC-type model galaxies, respectively.
In the respective pairs of lines, the upper and lower ones correspond to the models with the minimum $SB$ and the maximum $SB$ respectively. 
These lines are for the same seeing size and the same threshold level as for the observed data. 
Filled circles are galaxies classified as E or S0 in Dressler ($1980a$) and open circles are those classified as S.
Because of wide $SB$ range, the sample is divided into two panels in regard to $SB$. 
\label{fessep}}

\figcaption[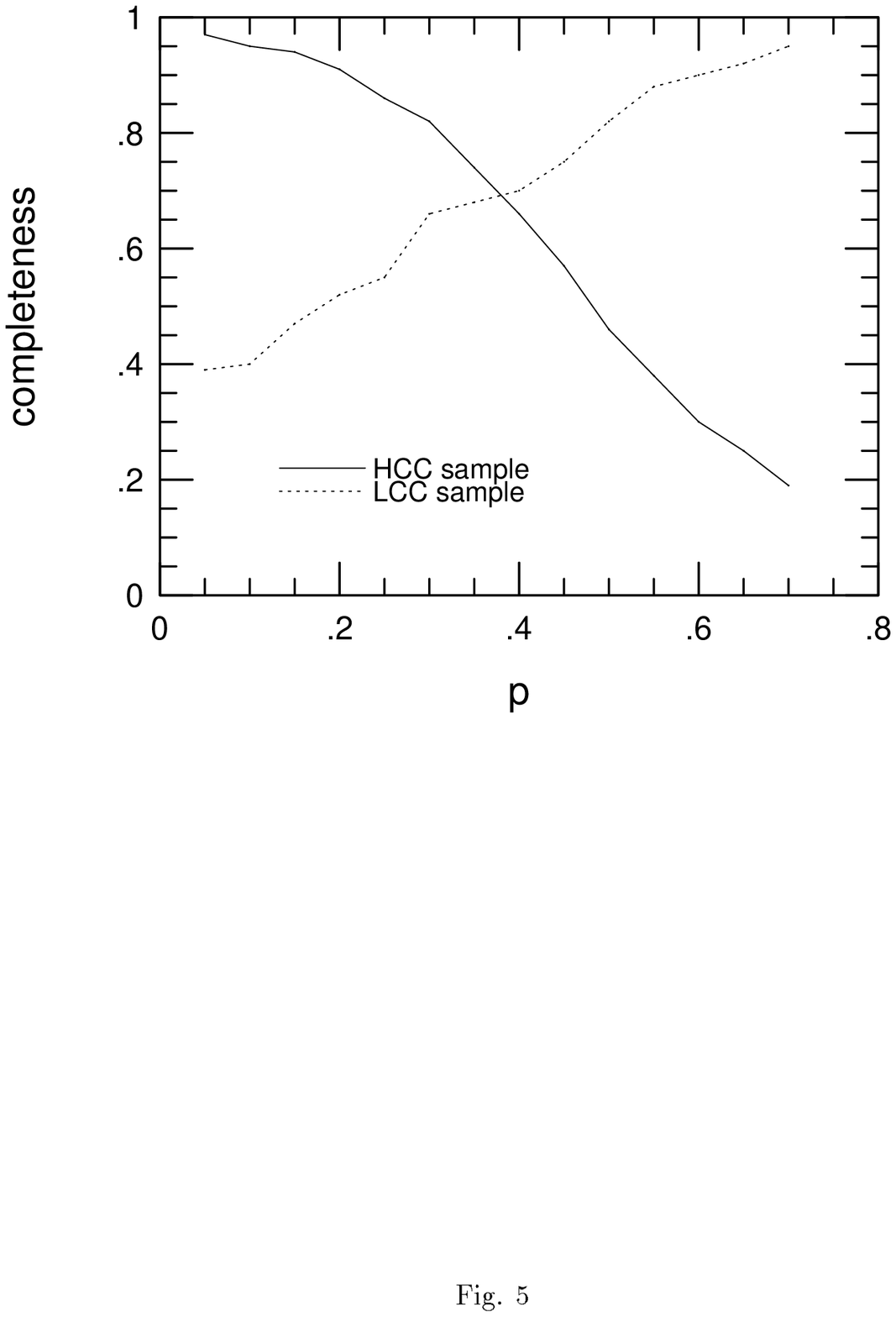]{Completeness of the HCC types (solid
line) and the LCC types (dotted line) as a function of $p$. We define the
^^ standard' sample with $p=0.35$ where the two curves intersect with each other. 
\label{fcomp}}

\figcaption[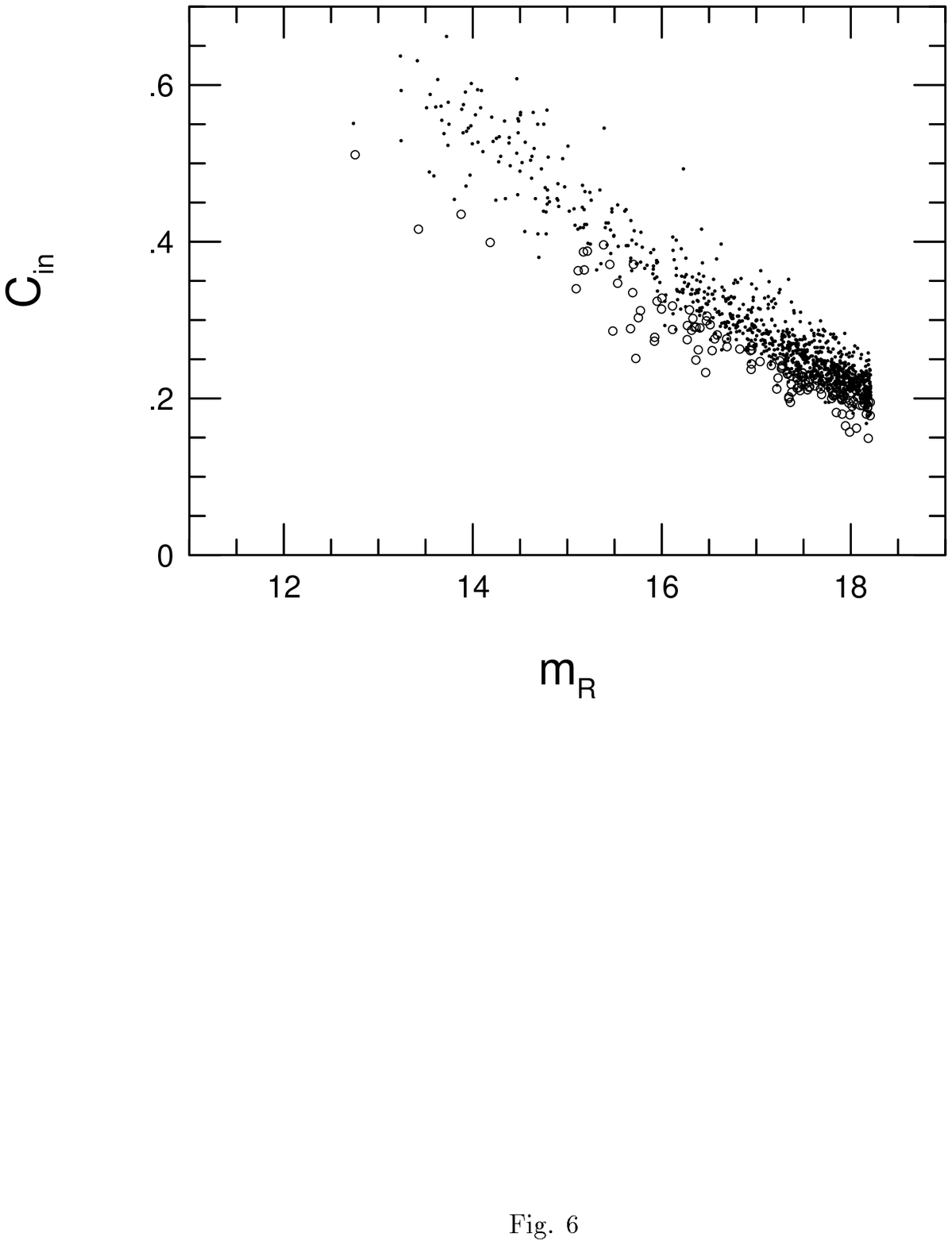]{Magnitude versus $\cin$ plot of the standard ($p=0.35$) sample.
Filled circles are galaxies classified as the HCC type, and open
circles are those classified as the LCC type.
\label{fes}}

\figcaption[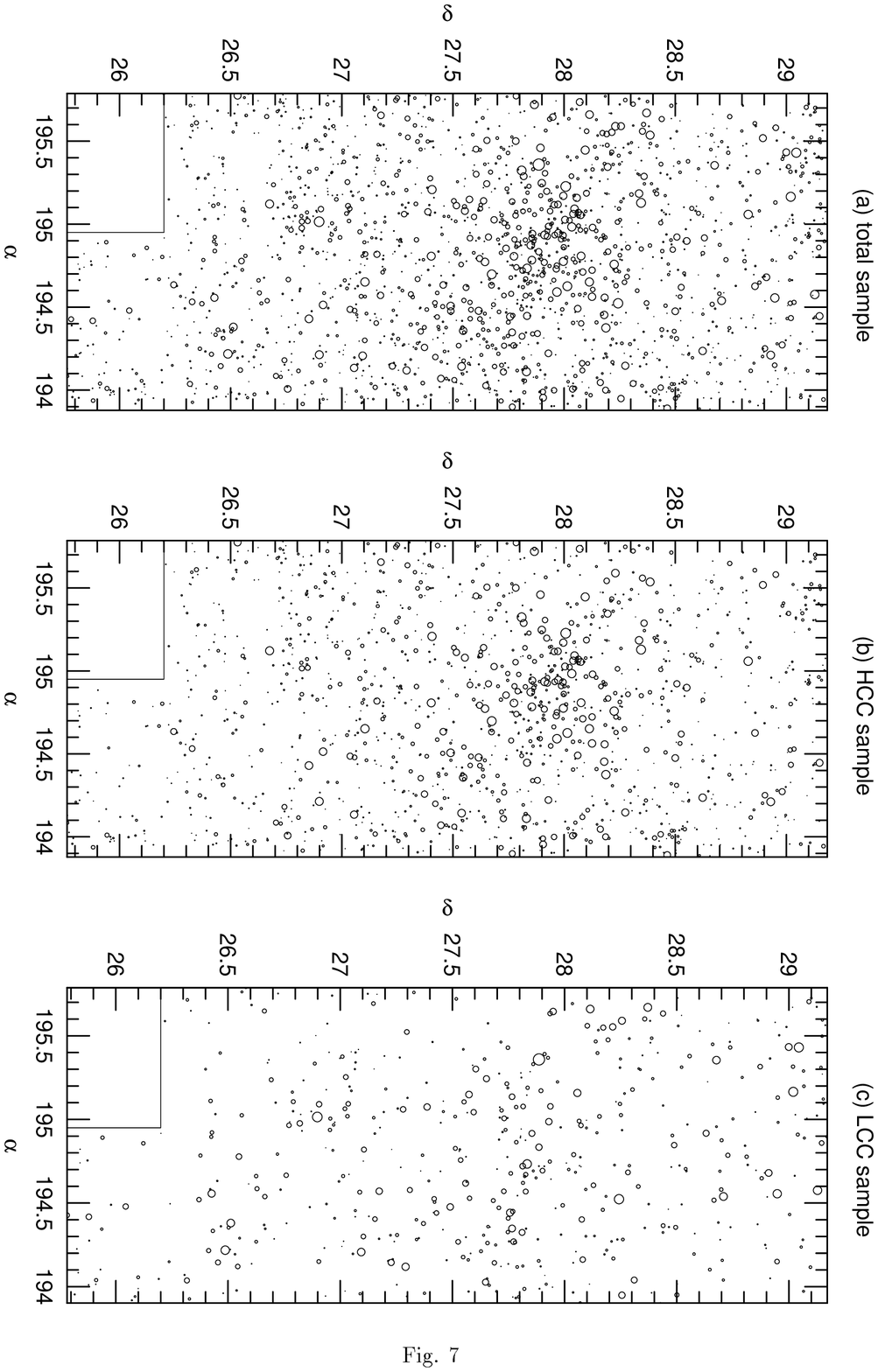]{Sky distribution of galaxies of (a) the total
sample, (b) the HCC sample, and (c) the LCC sample.
The size of a circle is proportional to the luminosity of the galaxy. 
\label{fsky}}

\figcaption[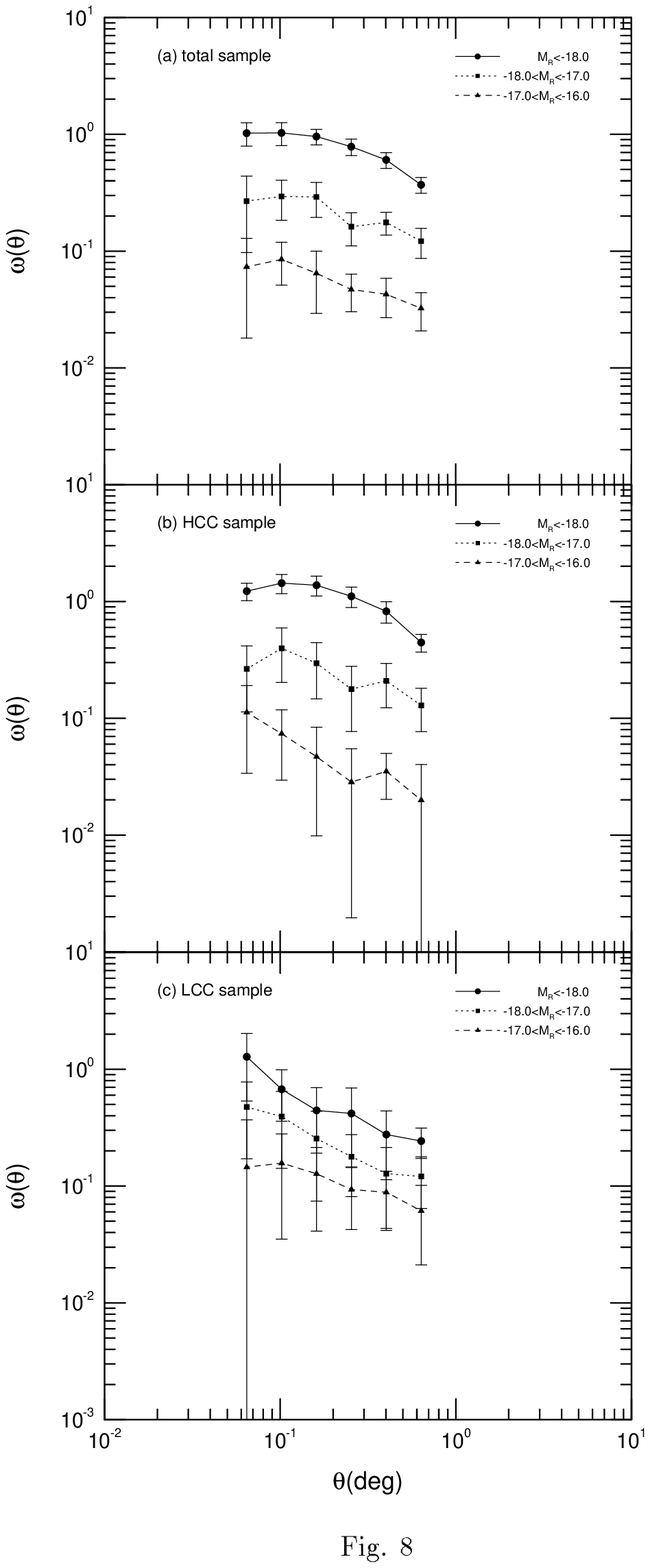]{The $\acf$s of subsamples with different
luminosity extracted from (a) the total sample, the (b) the HCC sample,
and (c) the LCC sample.
The solid line denotes the bright sample, the dotted line denotes the intermediate sample, and the dashed line denotes the faint sample.
\label{facfls}}

\figcaption[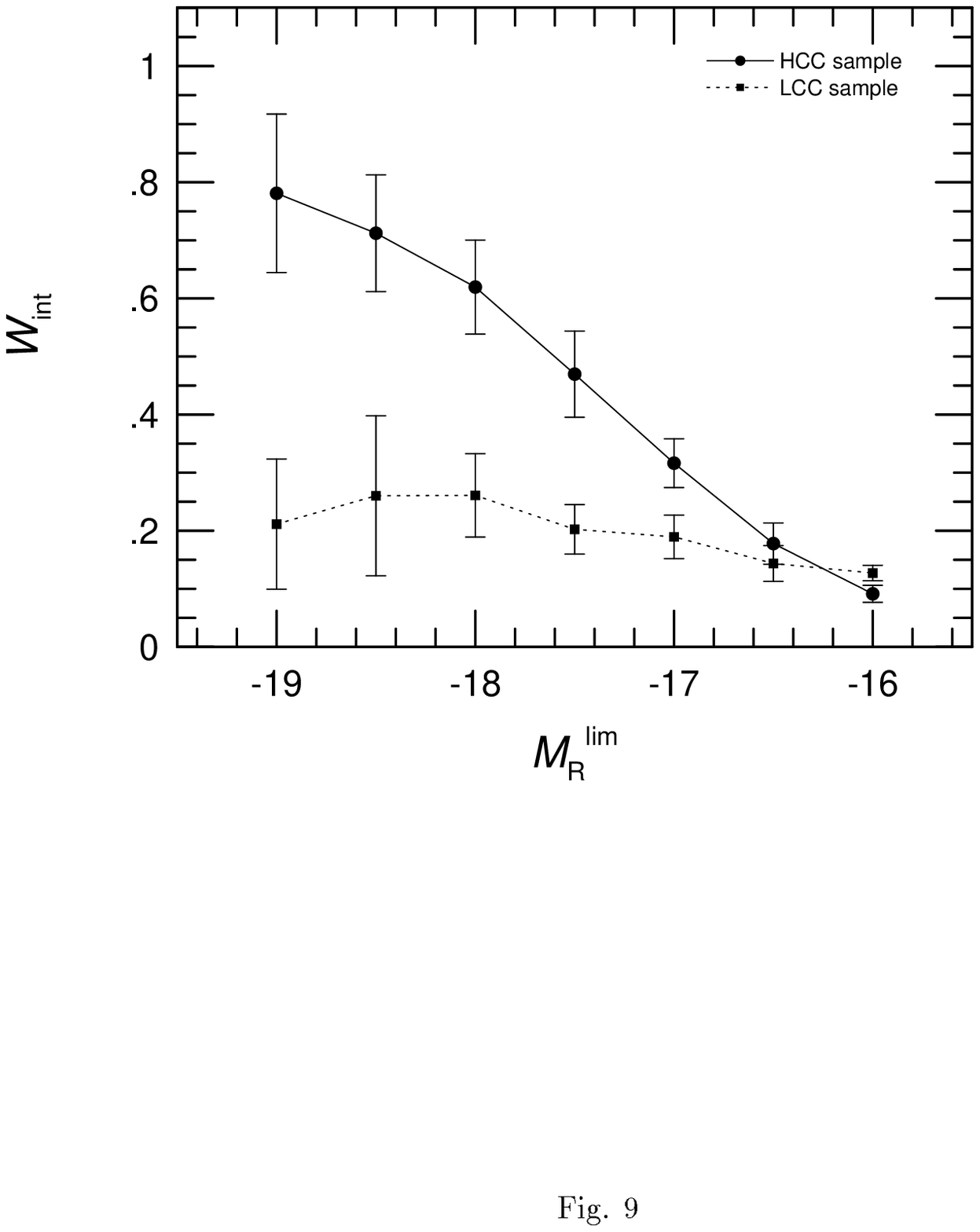]{$W_{int}$ as a function of limiting magnitude $M_R^{lim}$.
The solid line represents the HCC sample and the dotted line represents
the LCC sample.
The dashed line denotes the faintest magnitude limited by the contamination of field galaxy.
The error bars denote the fluctuations of $W_{int}$ estimated by BRM.
\label{fwintms}}

\figcaption[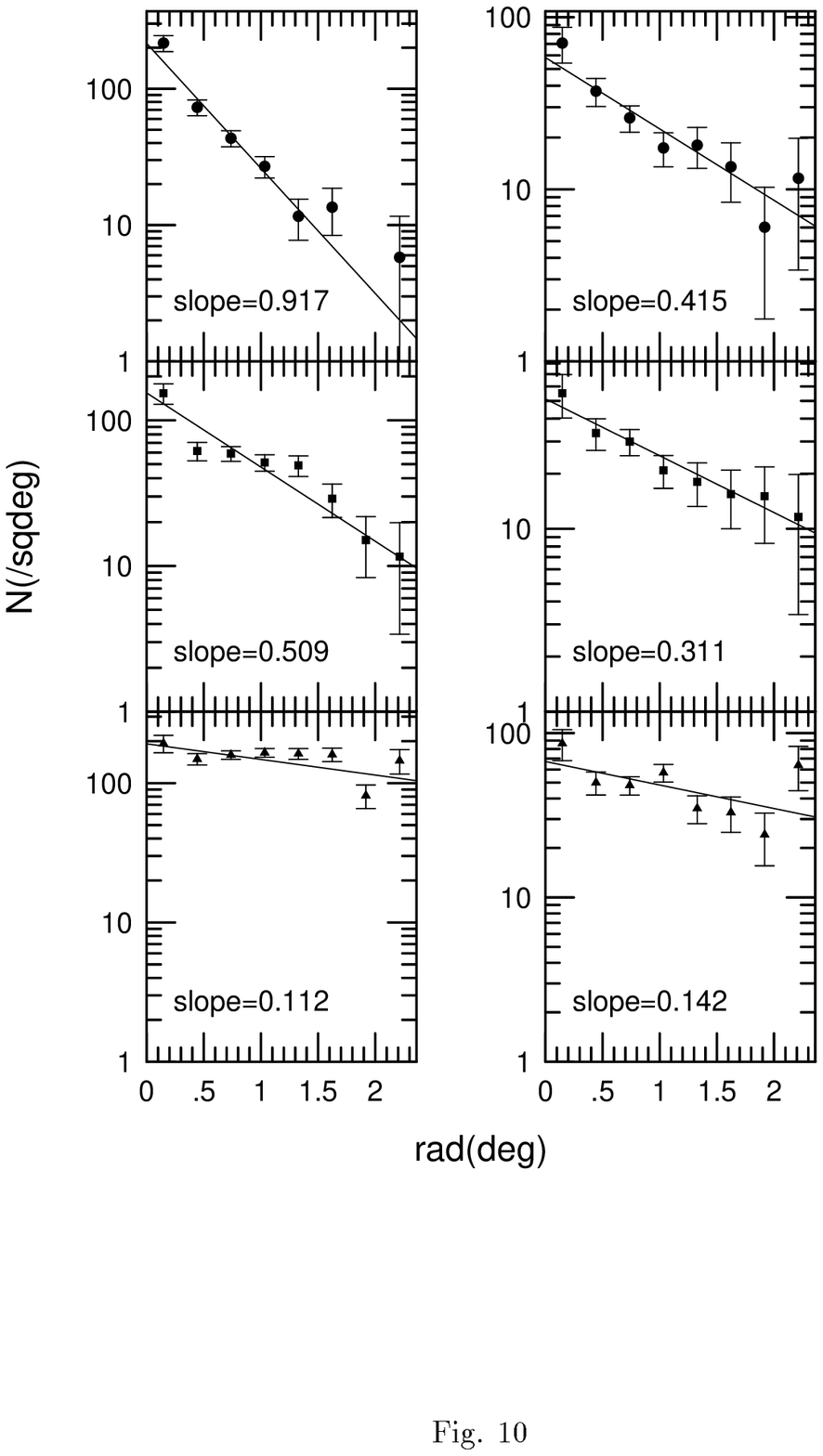]{Surface number density (/sqdeg)
of galaxies in the subsamples with different luminosities and morphologies as a
function of the distance from the cluster center. The left panels show the surface
number density for the HCC sample and the right panels show for the LCC
sample. From top to bottom: the bright sample, the intermediate sample, and
the faint sample. 
\label{fradp}}

\figcaption[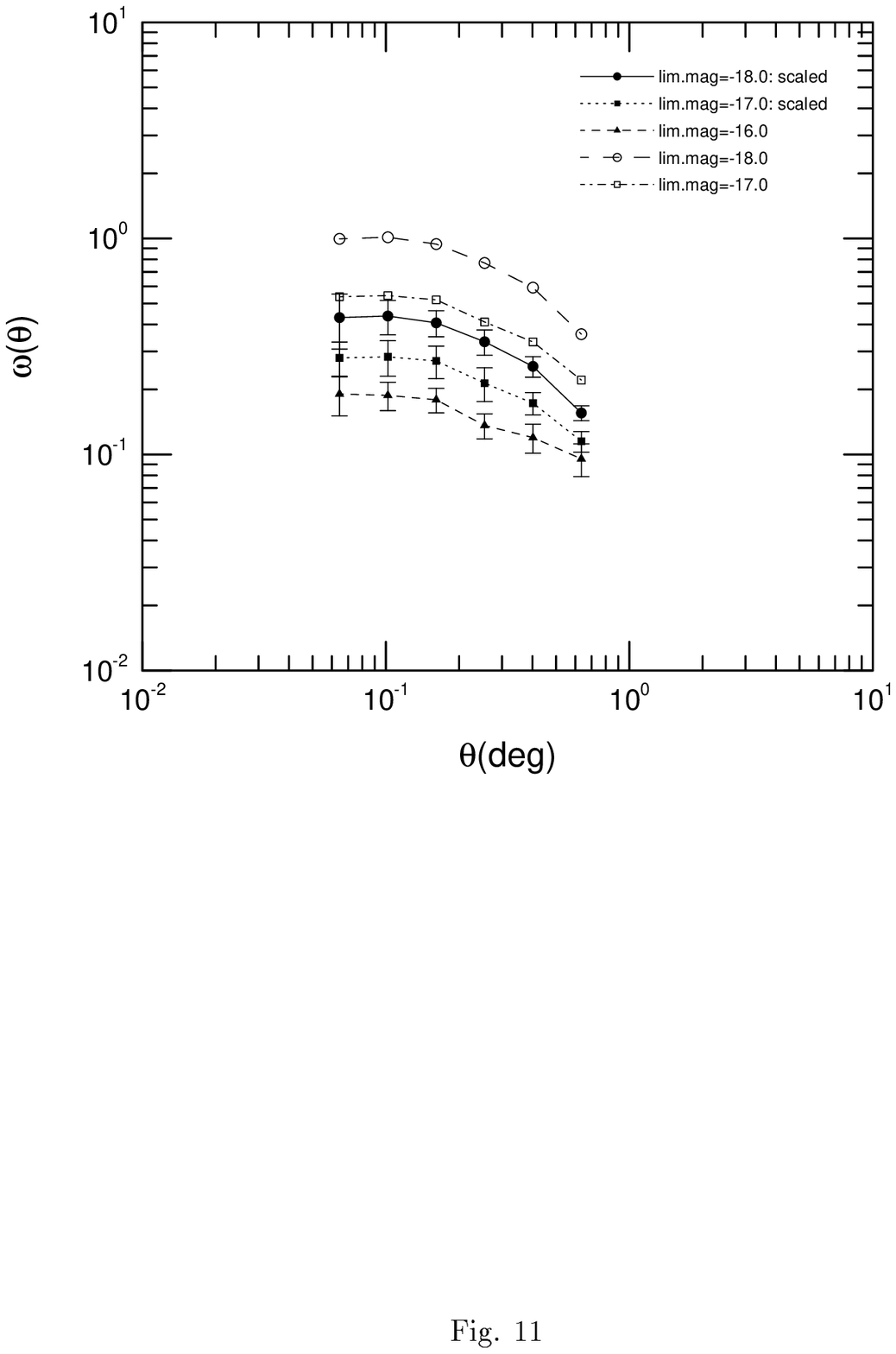]{Scaling test for the total sample. 
The sample with brighter limiting magnitude (filled circle and filled
square) is scaled to that with fainter limiting magnitude (filled
triangle) according to the values given in Table 3 (see the
text for details). The open symbols denote the $\acf$s for subsample
with brighter magnitude before scaling. 
\label{fstest}}

\figcaption[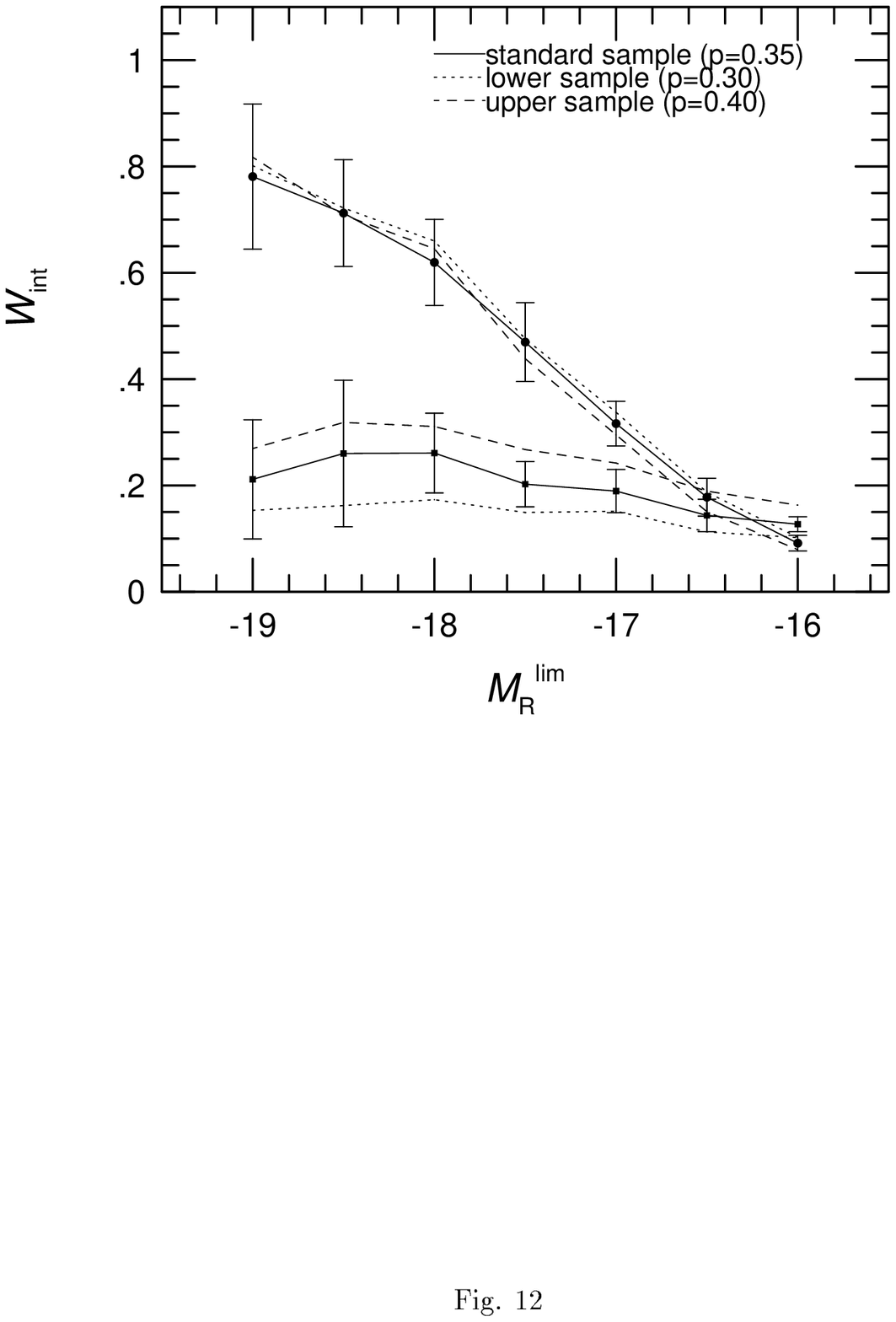]{$W_{int}$ as a function of
limiting magnitude when the lower and upper samples are used. The solid line
represents the standard sample (same as Figure 9)with one
sigma fluctuations of $W_{int}$ estimated by BRM.
The dotted line represents lower sample and the dashed line upper sample. 
\label{ferr}}

\clearpage 
 
\begin{deluxetable}{ccccc}
\tablecaption{The number of galaxies of our Coma cluster and SA57
samples and the expected contamination rate of field galaxies \label{tbgc}}
\tablehead{
\colhead{$m_R^{lim}$} & \colhead{$M_R^{lim}$} & \colhead{$N_{Coma}$\tablenotemark{a}} &
\colhead{$N_{SA57}$\tablenotemark{b}} & \colhead{contamination rate}\nl    
} 
\startdata
$15.73$ & $-18.5$ & $243$  & $68.5$ & $0.28$ \nl
$16.23$ & $-18.0$ & $338$  & $79.8$ & $0.24$ \nl
$16.73$ & $-17.5$ & $499$  & $125.$ & $0.25$ \nl
$17.23$ & $-17.0$ & $747$  & $228.$ & $0.31$ \nl
$17.73$ & $-16.5$ & $1172$ & $548.$ & $0.47$ \nl
$18.23$ & $-16.0$ & $1805$ & $890.$ & $0.49$ \nl
\enddata
\tablenotetext{a}{Number of galaxies in Coma sample (/5.25sqdeg)}
\tablenotetext{b}{Number of field galaxies(/5.25sqdeg) estimated from
SA57 sample}

\end{deluxetable}
 
\begin{deluxetable}{lrrcccccc}
\tablecaption{Fitting parameters of $\acf$ for total sample and
subsamples from different morphology and different luminosity \label{tacfls}}
\tablehead{
\colhead{morph.type} & \colhead{$M_R$} & \colhead{N} &
\colhead{$\gamma$} & \colhead{$\delta \gamma$} &
\colhead{$A_{\omega}$} & \colhead{$\delta A_{\omega}$} & \colhead{$W_{int}$} &
\colhead{$\delta W_{int}$} 
} 
\startdata
           &         -- $-18.0$ & $338$ & $-0.430$ & $0.090$ & $0.377$ & $0.066$ & $0.483$ & $0.064$\\
total      & $-18.0$ -- $-17.0$ & $409$ & $-0.379$ & $0.250$ & $0.113$ & $0.048$ & $0.139$ & $0.036$\\
           & $-17.0$ -- $-16.0$ & $1058$ & $-0.404$ & $0.251$ & $0.029$ & $0.015$ & $0.036$ & $0.013$\\
\hline
           &         -- $-18.0$ & $215$ & $-0.434$ & $0.106$ & $0.499$ & $0.113$ & $0.642$ & $0.106$\\
HCC        & $-18.0$ -- $-17.0$ & $281$ & $-0.377$ & $0.264$ & $0.126$ & $0.052$ & $0.154$ & $0.045$\\
           & $-17.0$ -- $-16.0$ & $817$ & $-0.711$ & $0.412$ & $0.014$ & $0.012$ & $0.024$ & $0.013$\\
\hline
           &         -- $-18.0$ & $123$ & $-0.690$ & $0.291$ & $0.156$ & $0.078$ & $0.257$ & $0.092$\\
LCC        & $-18.0$ -- $-17.0$ & $128$ & $-0.661$ & $0.393$ & $0.079$ & $0.051$ & $0.125$ & $0.058$\\
           & $-17.0$ -- $-16.0$ & $241$ & $-0.396$ & $0.544$ & $0.057$ & $0.027$ & $0.071$ & $0.023$\\  
\enddata

\end{deluxetable}

 
\begin{deluxetable}{cccccc}
\tablecaption{Parameters for scaling test \label{tst}}
\tablehead{
\colhead{$M_R^{lim}$} & \colhead{$N_{exp}$} & \colhead{$A$} &
\colhead{$f_{f}$} & \colhead{$f_{f}^2A$} & \colhead{$f_{c}^2\omega_{cc}$}
} 
\startdata
$-18.0$ & $15.2$  & $9.02\times10^{-2}$ & $0.24$ & $5.20\times10^{
-3}$ & $0.58\omega_{cc}$\\
$-17.0$ & $43.4$  & $5.22\times10^{-2}$ & $0.31$ & $5.02\times10^{
-3}$ & $0.48\omega_{cc}$\\
$-16.0$ & $169.5$ & $3.18\times10^{-2}$ & $0.49$ & $7.64\times10^{
-3}$ & $0.26\omega_{cc}$\\
\enddata

\end{deluxetable}


\clearpage


\begin{thebibliography}{}
\bibitem[Abraham et al.]{abr94} Abraham, R.G., Valdes, F., Yee, H.K.C., \&
van den Bergh, S. 1994, \apj, 432, 75
\bibitem[Barrow, Bhavsar, \& Sonoda 1984]{bbs84} Barrow, J.D., Bhavsar, S.P., \& Sonoda, D.H. 1984, \mnras, 210, 19
\bibitem[Bahcall \& Soneira 1983]{bs83} Bahcall, N.A., \& Soneira,
R.M. 1983, \apj, 270, 20 (BS83)
\bibitem[Binggeli, Tammann \& Sandage 1987]{bts87} Binggeli, B.,
Tammann, G.A. \& Sandage, A. 1987, \aj, 94, 251
\bibitem[Brainerd \& Villumsen 1992]{bv92} Brainerd, T.G., \& Villumsen, J. V. 1992, \apj, 400, 398
\bibitem[Briel, Henry \& B\"ohringer 1992]{bhb92} Briel, U.G., Henry,
J.P. \& B\"ohringer, H. 1992, \aap, 259, L31
\bibitem[Burstein \& Heiles 1984]{bh84} Burstein, D., \& Heiles, C. 1984, \apjs, 54, 33
\bibitem[Caldwell et al.\ 1993]{cal93} Caldwell, N., Rose, J.A., Sharples, R.M., Ellis, R.S., \& Bower, R.G., 1993, \aj, 106, 473
\bibitem[Capelato et al.\ 1980]{cap80} Capelato, H.G., Gerbal, D., Mathez, G., Mazure, A., Salvadole-Sole, E., \& Sol, H. 1980, \apj, 241, 521
\bibitem[Davis \& Geller 1976]{dg76} Davis, M. \& Geller, M.J. 1976, \apj, 208, 13
\bibitem[Davis et al.\ 1985]{dav85} Davis, M., Efstathiou, G., Frenk, C.S., White, S.D.M. 1985, \apj, 292, 371
\bibitem[Davis et al. 1988]{dav88} Davis, M., Meiksin, A., Strauss, A., Nicolaci da Costa, L., \& Yahil, A. 1988, \apj, 333, L9
\bibitem[de Lapparent, Geller, \& Huchra 1988]{lgh88} de Lapparent, V., Geller, M.J., \& Huchra, J.P. 1988, \apj, 332, 44
\bibitem[Diaconis \& Efron 1983]{de83} Diaconis, P., \& Efron, B. 1983, Sci.Am., 1993, 96
\bibitem[Doi, Fukugita, \& Okamura 1993]{dfo93} Doi, M., Fukugita, M., \& Okamura, S. 1993, \mnras, 264, 832
\bibitem[Doi, Fukugita, \& Okamura 1995]{dfo95} Doi, M., Fukugita, M.,
\& Okamura, S. 1995, \apjs, 97, 59
\bibitem[Dom\'inguez-Tenreiro \& Pozo-Sanz 1988]{dp88} Dom\'inguez-Tenreiro, R., \& Pozo-Sanz, R.D. 1988, \apj, 324, 677 
\bibitem[Dressler 1980a]{dre80a} Dressler, A. 1980a, \apjs, 42, 565
\bibitem[Dressler 1980b]{dre80b} Dressler, A. 1980b, \apj, 236, 351
\bibitem[Eder et al. 1989]{ed89} Eder, J.A., Schombert, J.M., Dekel,
A., \& Oemler, A.Jr. 1989, \apj, 340, 29 
\bibitem[Efstathiou, Ellis, \& Peterson 1988]{eep88} Efstathiou, G., Ellis, R.S., \& Peterson, B.A. 1988, MNRAS, 232, 431 
\bibitem[Einasto 1991]{ein91} Einasto, M. 1991, \mnras, 250, 802
\bibitem[Ferguson \& Binggeli 1994]{fb94} Ferguson, H.C., \& Bingelli, B. 1994, \aapr, 6, 67
\bibitem[Ferguson \& Sandage 1989]{fs89} Ferguson, H.C., \& Sandage, A., 1989, \apj, 346, L53
\bibitem[Forbes \& Dawson 1986]{fd86} Forbes, D., \& Dawson, P.C. 1986, \pasp, 98, 102
\bibitem[Garili et al. 1992]{gar92} Garili, B., Bottini, D., Maccagni, D., Vettolani, G., \& Maccacaro, T. 1992, \aj, 104, 1290
\bibitem[Groth 1986]{gro86} Groth, E.J. 1986, \aj, 91, 1244
\bibitem[Groth \& Peebles 1977]{gp77} Groth, E.J., \& Peebles,
P.J.E. 1977, \apj, 217, 385  
\bibitem[Hamilton 1988]{ham88} Hamilton, A.J.S. 1988, \apj, 331, L59
\bibitem[Kaiser 1984]{kai84} Kaiser, N. 1984, \apj, 284, L9
\bibitem[Kashikawa et al. 1995a]{kas95a} Kashikawa, N., Sekiguchi, M.,
Yagi, M., Yasuda, N., Shimasaku, K., Doi, M., \& Okamura, S., 1995a, in "Scientific and Engineering Frontiers for 8-10m Telescopes", M. Iye and T. Nishimura eds., Universal Academy Press, p.105
\bibitem[Kashikawa et al. 1995b]{kas95b} Kashikawa, N., Shimasaku, K.,
Yagi, M., Yasuda, N., Doi, M., Okamura, S., \& Sekiguchi, M., 1995b, \apj, 452, L99 
\bibitem[Koo \& Szalay 1984]{ks84} Koo, D.C., \& Szalay, A.S. 1984, \apj, 282, 390
\bibitem[Loveday et al. 1995]{lov95} Loveday, J., Maddox, S.J., Efstathiou, G., \& Peterson,
B.A. 1995, \apj, 442, 457
\bibitem[Lasker et al. 1990]{las90} Lasker, B.M., Sturch, C.R.,
McLean, B.J., Russell, J.L., Jenker, H., \& Shara, M.M., \aj, 99, 2019
\bibitem[Maurogordato \& Lachieze-Rey 1991]{ml91} Maurogordato, S. \& Lachieze-Rey, M. 1991, \apj, 369, 30
\bibitem[Mo, Jing, \& B\"orner 1992]{mjb92} Mo, H.J., Jing, Y.P., \& B\"orner, G. 1992, \apj, 392, 452
Dressler, A., \& Oemler, Jr.A. 1996, \nat, 379, 613
\bibitem[Neuschaefer, Windhorst, \& Dressler 1991]{nwd91} Neuschaefer, L.W., Windhorst, R.A., \& Dressler, A. 1991, \apj, 382, 32
\bibitem[Oemler 1974]{oem74} Oemler, A.Jr. 1974, \apj, 194, 1
\bibitem[Okamura, Kodaira, \& Watanabe 1984]{okw84} Okamura, S.,
Kodaira, K., \& Watanabe, M. 1984, \apj, 280, 7
\bibitem[Okamura et al. 1997]{oka97} Okamura, S., Doi, M., Kashikawa,
N., Kawasaki, W., Komiyama, Y., Sekiguchi, M., Shimasaku, K., Yagi,
M., \& Yasuda, N. 1996, Proc. IAU Symposium No. 179: "New Horizons from Multi-Wavelength Sky Surveys",  in press
\bibitem[Peebles 1980]{pee80} Peebles, P.J.E. 1980, in The Large-Scale Structure of the Universe (Princeton: Princeton Univ. Press) 
\bibitem[Phillips \& Shanks 1987]{ps87} Phillips, S. \& Shanks, T. 1987, \mnras, 229, 621
\bibitem[Pritchet \& Infante 1992]{pi92} Pritchet, C.J., \& Infante, L. 1992, \apj, 399, L35
\bibitem[Quintana 1979]{qui79} Quintana, H. 1979, \aj, 84, 15
\bibitem[Salzer, Hanson, \& Gavazzi 1990]{shg90} Salzer, J.J., Hanson, M.M., \& Gavazzi, G. 1990, \apj, 353, 39
\bibitem[Santiago \& da Costa 1990]{sd90} Santiago, B.X., \& da Costa, L.N. 1990, \apj, 362, 386
\bibitem[Sekiguchi et al. 1992]{sek92} Sekiguchi, M., Iwashita, H., Doi, M., Kashikawa, N., \& Okamura, S. 1992, \pasp, 104, 744
\bibitem[Shanks et al. 1980]{sha80} Shanks, T., Fong, R., Ellis, R.S., \& MacGillivray, H.T. 1980, \mnras, 192, 209
\bibitem[Thompson \& Gregory 1993]{tg93} Thompson, L.A. 1993, \aj, 106, 2197
\bibitem[Tully 1988]{tul88} Tully, R.B. 1988, \aj, 96, 173
\bibitem[Valls-Gebaud, Alimi, \& Blanchard 1989]{vab89} Valls-Gebaud, D., Alimi, J., \& Blanchard, A. 1989, \nat, 341, 215
\bibitem[van den Bergh 1986]{van86} van den Bergh, S. 1986, \aj, 91, 271
\bibitem[Villumsen, Freudling, \& da Costa 1997]{vfd97} Villumsen, J.,
Freudling, W., \& da Costa, L. 1997, \apj, 481, 578
\bibitem[White et al. 1987]{wh87} White, S.D.M., Davis, M., Efstathiou, G., \& Frenk, C.S. 1987, \nat, 330, 451
\bibitem[Yagi et al. 1998]{y98} Yagi, M. et al. 1998, in preparation 
\bibitem[Yamagata \& Yoshii 1992]{yy92} Yamagata, T. \& Yoshii, Y. 1992, \aj, 103, 117 (YY92)
\bibitem[Yoshii \& Takahara 1988]{yt88} Yoshii, Y. \& Takahara, F. 1988, \apj, 326, 1
\bibitem[Yoshii 1993]{yos93} Yoshii, Y. 1993, \apj, 403, 552
\bibitem[Zabludoff, Geller, \& Huchra, 1993]{zgh93} Zabludoff, A.I., Geller, M.J., Huchra, J.P. 1993, \aj, 106, 1273
\end{thebibliography}
\end{document}